\documentclass[runningheads]{llncs}
\usepackage[misc]{ifsym}
\usepackage{graphicx}
\usepackage{upgreek}
\usepackage{amssymb}
\usepackage{dsfont}
\usepackage{xspace}
\usepackage{booktabs}
\newcommand{\ra}[1]{\renewcommand{\arraystretch}{#1}}
\usepackage{graphicx}
\usepackage[export]{adjustbox}
\usepackage{amsfonts}
\usepackage{tikz}
\usepackage{bm}
\usepackage{multirow}
\usepackage{diagbox}
\usepackage{rotating}
\usepackage{listings}
\usepackage{mathtools}
\usepackage{adjustbox}
\usepackage{multirow}
\usepackage{graphics}
\usepackage{capt-of}
\usepackage{caption}
\usepackage{subcaption}
\usepackage[all=normal,tracking=tight,paragraphs=tight]{savetrees}
\graphicspath{ {../Figure/} }
\usepackage{pifont}
\newcommand{\xmark}{\ding{55}}%

\newcommand{\m}{s}

\newcommand{\Prob}{\textbf{{Pr}}}
\newcommand{\Expec}{\textbf{{E}}}
\newcommand{\til}{~}
\newcommand{\HB}{HyperANF}
\newcommand{\mhse}{MHSE}

\newcommand{\bigO}{\mathcal{O}}
\newcommand{\Bool}{{\sc{propagate-p}}}
\newcommand{\BoolSE}{{\sc{propagate-s}}}
\newcommand{\Propagate}{{\sc{propagate}}}
\newcommand{\PropagateS}{{\sc{prop.}}}

\usepackage[linesnumbered,ruled,vlined,boxed]{algorithm2e}
\newlength{\commentWidth}
\let\oldnl\nl
\newcommand{\nonl}{\renewcommand{\nl}{\let\nl\oldnl}}
\DontPrintSemicolon
\SetKwInOut{Input}{Input}\SetKwInOut{Output}{Output}
\SetKwRepeat{Do}{do}{while}

\newcommand{\sigg}[1]{\texttt{Sig}(#1)}
\newcommand{\Sigg}[2]{\texttt{Sig}_{#2}(#1)}

\begin{document}

	\title{\Propagate: a seed propagation framework to compute Distance-based metrics on Very Large Graphs}
	
	\author{Giambattista Amati\inst{1} \and
		Antonio Cruciani\inst{2} (\Letter) \and
		Daniele Pasquini\inst{3}\and\\ Paola Vocca\inst{3} \and Simone Angelini\inst{1}  }
	\authorrunning{Amati et al.}
 \titlerunning{Computing Distance-based metrics on Very Large Graphs}
	\institute{Fondazione Ugo Bordoni
		Rome, Italy\\ \email{\{gba,sangelini\}@fub.it}\and
		Gran Sasso Science Institute
		L’Aquila, Italy\\ \email{antonio.cruciani@gssi.it}\and University of Rome ``Tor Vergata''
		Rome, Italy\\
		 \email{\{daniele.pasquni,paola.vocca\}@uniroma2.it}\\
		}
	\toctitle{\Propagate: a seed propagation framework to compute Distance-based metrics on Very Large Graphs}
	\tocauthor{Giambattista~Amati, Antonio~Cruciani, Daniele~Pasquini, Paola~Vocca, Simone~Angelini}
	
	\maketitle
	\begin{abstract}
We propose \Propagate, a fast approximation framework to estimate distance-based metrics on very large graphs such as: the (effective) diameter  or the average distance within a small error. The framework assigns seeds to nodes and propagates them in a BFS-like fashion, computing the neighbors set until we obtain  either the whole vertex set (for computing the diameter) or a given percentage of vertices (for the effective diameter). At each iteration, we derive  compressed Boolean representations of the neighborhood sets discovered so far. The \Propagate\til framework yields two algorithms: \Bool, which propagates all the $\m$ seeds in parallel, and \BoolSE\til  which propagates the seeds   sequentially. 
For each node, the compressed representation of the \Bool\til  algorithm requires $\m$ bits while \BoolSE\til  $1$ bit only. Both algorithms  compute the average distance, the effective diameter, the diameter, and the connectivity rate (a measure of the the sparseness degree of the  transitive closure graph)  within a small error with high probability: for any $\varepsilon>0$ and using $\m=\Theta\left(\frac{\log n}{\varepsilon^2}\right)$ sample nodes, the error for the average distance is bounded by $\xi  = \frac{\varepsilon \Delta}{\alpha}$; the errors for the effective diameter and the diameter are bounded by $\xi = \frac{\varepsilon}{\alpha}$; and the error for the connectivity rate is bounded by $\varepsilon$ where $\Delta$ is the diameter and $\alpha$ is the connectivity rate. The time complexity of our approaches is $\bigO(\Delta\cdot m)$ for \Bool~and $\bigO\left(\frac{\log n}{\varepsilon^2}\cdot \Delta \cdot m\right)$ for \BoolSE, where $m$ is  the number of edges of the graph and $\Delta$ is the diameter.The experimental results show that  the \Propagate\til framework improves the current state of the art in accuracy, speed, and space.  Moreover, we experimentally show that \Propagate\til is also very efficient for solving the All Pair Shortest Path problem in very large graphs.

\end{abstract}
	
\section{Introduction}\label{sec:introduction}

%\add{Distance-based metrics are notions for evaluating a wide range of properties on graphs, used in network analysis and graph theory. These, provide insight about how far nodes can be the one from each other. Distance-based metrics can be used to compute notions of ``importance'' in networks i.e., centrality measures, cohesion among nodes  .

%	A key quantity from which distance-based metrics can be easily obtained is the \emph{neighborhood function} $N(r)$, that for each $r\in \mathbb{N}$ it returns the number of pairs of nodes $(u,v)$ such that $v$ is reachable from $u$ in at most $r$ hops. From the neighborhood function, distance-based metrics can be computed} 

The fast computation of distances between pairs of nodes in a graph  is a fundamental task in network applications. Distance-based metrics are  also used to compute different notions of centrality for nodes or edges that can be used to detect communities in very large graphs,  as proposed  by Girvan and Newman~\cite{Girvan7821}  or  Fortunato et al.~\cite{Fortunato:2004}.
The \emph{diameter},   i.e.  the maximum distance between all  reachable pairs in a graph, is an important parameter for analyzing  graphs that, for example, change over the time~\cite{Kumar:2006}, or real-world graphs as the web  and social network graphs, which have small diameters~\cite{Kleinberg:2000:SPA} that shrink as  they grow~\cite{Leskovec:2005:GOT}. 
The fastest  exact algorithm for computing the diameter of \emph{sparse graphs}  is based on solving the  All-Pairs Shortest Paths (APSP) problem which, for unweighted graphs, can be computed by executing a Breadth-First Search (BFS) for each vertex, with  a time complexity of $\Omega(mn)$, where $n$ is the number of nodes  and $m$ the number of edges.
%The fastest  exact algorithm to compute the diameter of \emph{sparse graphs}  is based on solving the  All-Pairs Shortest Paths (APSP) problem~\cite{Cormen:2009:IAT}, that has a running time of $O(mn +n^2\log n)$, where $n$ is the number of nodes  and $m$ the number of edges.  
%For some classes of graphs (undirected, or unweighted, or acyclic) the time complexity is $O(m n)$ \cite{Thorup:1999,,}.
%and for 
For \emph{dense}  graphs, the best algorithm is based on matrix multiplication~\cite{DBLP:journals/jacm/CyganGS15}, which can be performed in time of $\tilde{\bigO}(n^{\omega})$, where $\omega<2.38$~\cite{COPPERSMITH1990251,Williams12}.
However, its well known that computing the diameter of a graph with $m$ edges requires $m^{2-o(1)}$ time under the Strong Exponential Time Hypothesis (SETH), which can be prohibitive for very large graphs \cite{Abboud_2014,Dalirrooyfard_2021}, so \emph{efficient approximation algorithms} for diameter are highly desirable. A trivial $2$-approximation algorithm for the exact diameter in undirected graphs can be computed in $\bigO(m+n)$ time by means of a BFS-visit starting from an arbitrary node. %, while  the exact algorithm  time complexity  is $O(mn)$~\cite{Thorup:1999}. 
A ${3}/{2}$-approximation algorithm was first presented by Aingworth \emph{et al.}~\cite{doi:10.1137/S0097539796303421} with a time complexity of $\tilde{\bigO}(m\sqrt{n}+n^2)$,  further improved to $\tilde{\bigO}(m\sqrt{n})$~\cite{DBLP:RodittyW13},  and, with the same approximation ratio,  to  $\tilde{\bigO}(m^{3/2})$ or $o(n^2)$, depending on the degree of sparsity of the graph~\cite{Chechik:2014:BAA}. If  a graph is  weakly connected, experiments with real-world graph data sets show that heuristics may decrease the average  running time of the diameter computation~\cite{DBLP:journals/tcs/BorassiCHKMT15}. The computation of the exact diameter  is however  susceptible to outliers. For this reason,  it is preferable  to use more robust metrics, such as  the \emph{effective diameter}, which is  defined as the a percentile distance between nodes (e.g. $90^{th}$), i.e.  the maximum  distance that allows to connect that percentage of all reachable pairs~\cite{Palmer2001,tauro2001simple}.
For large real graphs, even the exact computation of the  effective diameter  remains prohibitive since  possible approaches  are still based either on solving  APSP or on computing a transitive closure.
Also,  some diameter approximation algorithms~\cite{DBLP:journals/tcs/BorassiCHKMT15,CPPU_2020} cannot be used to compute the effective diameter, that is because they are based on the computation of the greatest distances from the nodes that do not necessarily pass through all reachable pairs \cite{DBLP:journals/tcs/BorassiCHKMT15} or on merging the diameters independently computed on smaller subgraphs \cite{CPPU_2020}. An alternative approach  is to compute the \emph{neighborhood function}  to derive  distance metrics. A \emph{neighborhood} $N(u,r)$ is the set of all nodes reachable from the node $u$ by a path of length at most $r$. $N(u,r)$ is also known as the \emph{ball} of center $u$ and radius $r$. The most efficient algorithms for approximating the effective diameter are based on the estimate of the size of neighborhoods. For example, ANF~\cite{palmer2002anf} is based on BFS and the use of Flajolet-Martin (FM) probabilistic counters~\cite{FlajoletMartin:1985},  and  HyperANF~\cite{Boldi:2011:HAN} is based on the same approach as ANF but with the use of HyperLogLog as probabilistic counter~\cite{DurandFlajolet2003}. 
Cohen~\cite{Cohen:2014:ASR} uses an approach  based on  a non-probabilistic counter, that uses $k$ hash functions  on neighborhoods  by keeping only the minimum hash value (MinHash) for each hash function ($k$-mins sketches). When the hashing values are in the unit interval $[0,1]$, then it is possible to estimate $ |N (u, r )|$ by means of the unbiased cardinality estimator  $\frac{1}{\textrm{MinHash}\left(N(u, r )\right)}$, with the standard error a function  of $k$.
The  \mhse\til framework\til\cite{BIGDATA2017}, instead, uses the  MinHash approach to derive  dense representations (\emph{signatures}) of large and sparse graphs that preserve similarity and  thus providing an approximation of the size of the neighborhood  of a node using the Jaccard similarity. 
%A similar approach, with a more general definition of probabilistic counters, is also used in~\cite{Cohen:2014:ASR}.
ANF, HyperANF and \mhse\til are grounded on the observation that  the size of $N(u, r)$ is sufficient to estimate the distance-based metrics.

\subsubsection*{Our contributions.}

%In all practical settings and actual graph sizes,  SE-MHSE is better than HyperANF in terms of space occupancy (see Fig. \ref{MHSE-HyperANF}). 
We propose a framework to estimate the distance-based metrics on graphs based on a mixed approach: \emph{sampling} and   \emph{counting}. The core idea of our approach is to consider a small set of $s$ seed nodes and to count the nodes that can be reached by at least one of these seeds, that is, the size of the neighbourhood set at distance $d$.
%We propose a \emph{sampling-based} framework to estimate the average distance and  the effective diameter of a given graph $G=(V,E)$,  considering a small set of $s$  sample nodes and  counting the neighbors that contain at least one of these nodes.
%Our \emph{sampling} method consists in counting the number of neighbors that contain nodes of this sample. 
 We define two implementations of our framework: \Bool, and \BoolSE. The time complexity of our approaches is, respectively, $\bigO(\Delta\cdot m)$, and $\bigO(s\cdot \Delta\cdot m)$, while the space complexity is $\bigO\left(s \cdot n +m \right)$, and $\bigO(n+m)$. 
 %This approach leads to \Bool\til algorithm  (Algorithm \til\ref{MHSE-Bool}), which has time complexity $\bigO\left(m\cdot\Delta\cdot\m\right)$ and uses space,  using a bit-like array to keep track of the information. Furthermore, since the algorithm only counts the neighbors that contain nodes from the sample, we can  process each sample node $s_i$, $1\leq i\leq s$, one at time, and derive a linear space  %(see Theorem  \ref{th:Algo2}) 
%This approach leads to \Bool\til algorithm  (Algorithm \til\ref{MHSE-Bool}), which has time complexity $\bigO\left(m\cdot\Delta\cdot\m\right)$ and uses space,  using a bit-like array to keep track of the information. Furthermore, since the algorithm only counts the neighbors that contain nodes from the sample, we can  process each sample node $s_i$, $1\leq i\leq s$, one at time, and derive a linear space  %(see Theorem  \ref{th:Algo2}) 
%version of the algorithm, namely \BoolSE\til algorithm (see Algorithm \til\ref{SE-MHSE}). 
We provide an estimate on the sample size needed to achieve a good estimate of the distance metrics up to a small error bound. More precisely,  we prove that $s=\Theta(\frac{\log n}{\varepsilon^2})$ sample nodes are sufficient to estimate, with probability at least $1-\frac{2}{n^2}$: (1) the average distance with the error bounded by $\varepsilon \frac{\Delta}{\alpha}$; (2) both the effective diameter and the diameter with the error bounded by  $\frac{\varepsilon} {\alpha}$; and, (3) the connectivity rate $\alpha$ with error bounded by $\varepsilon$, where $\alpha$ be  the \emph{connectivity rate} of the network (see Section \ref{sec_preliminaris} for the formal definition). 
%From this result we derive that  $\bigO\left(m \cdot\Delta\cdot\frac{\log n}{\varepsilon^2} \right)$ time is sufficient to obtain a  good approximation with high probability (w.h.p.). \add{We conduct our study by comparing our novel framework to the state-of-the-art algorithms for approximating distance-based metrics: (1) \HB~\cite{Boldi:2011:HAN}, namely the fastest available approximation algorithm able to approximate distance-based metrics on huge graphs; (2) the more accurate \textsc{MHSE}~\cite{BIGDATA2017}; and, (3) the well known algorithm \textsc{rand-BFS}~\cite{Eppstein:2001:FAC} by Eppstein et al. Our evaluation indicates that \Propagate~ consistently outperforms its competitors in terms of accuracy, execution time, and scalability. Next, we show that our \BoolSE's implementation its way better than \texttt{WebGraph}'s APSP algorithm \cite{BoVWFI} to compute the \emph{exact} distance-based metrics. }
It is important to underline that both the algorithms admit a straightforward and simple  implementation in a fully distributed and parallel setting. 

In Section~\ref{Sec:relatedwork}, we give an overview of relevant results on approximation algorithms of distance based metrics.
%In Section~\ref{Sec:MinHash}, we introduce the MinHash Signature Estimation framework. 
In Section~\ref{sec_preliminaris} we provide some basic preliminaries to understand our work.
In Section\til\ref{Sec:algorithm}, we describe the core idea behind our novel framework, then we introduce the new algorithms \Bool\til and \BoolSE, and provide an unbiased error bound for the computation of the effective diameter, the diameter, the average distance, and the connectivity rate.
In Section~\ref{Sec:tool}, we compare our framework with the state-of-the-art algorithms for approximating the distance-based metrics. Finally, in Section~\ref{Sec:conclusion}, we conclude and present future research directions.

\section{Related Works}
\label{Sec:relatedwork}

%\nota{Some recent works by Ceccarello et al. on distributed diameter approximation should be referenced.}
The literature on approximating distance-based metrics being vast, we restrict our attention to approaches that are closest to ours. We, thus, particularly focus on \emph{sampling} and \emph{probabilistic }techniques. 
%There are two main approaches to estimate the \add{distance-based metrics} on graphs: \emph{i)} \emph{sampling}: that is randomly drawing a sub-graph or more generally a set of nodes by which to exactly compute the measures of interest; and, \emph{ii)} \emph{probabilistic}: based on computing the  neighborhood sets through a suitably graph traversal and estimating their sizes using a probabilistic counter. %In the following, we  discuss the  methods in more depth.

\paragraph{Estimating Diameter by Sampling\label{SEC::SAMPLING}.} 
There are three main   questions  to be addressed  when sampling from large graphs~\cite{Leskovec:2006}: how to  sample nodes and  edges, how to set a  good sample size, how to  evaluate the goodness of the sample, as well as the goodness of the chosen sampling method. In the case of undirected and connected graphs, the centrality of nodes can be estimated  by sampling  only  $\bigO(\frac{\log n}{\varepsilon^2})$ nodes and compute all the distances to all other nodes, with an error of $\varepsilon\Delta$, where $\Delta$ is  the graph diameter~\cite{Eppstein:2001:FAC,Crescenzi_2011}, thus reducing the time complexity to $\bigO(\frac{\log n}{\varepsilon^2}\left (n\log n +m\right))$. %In the frequent case of graphs changing over time,
%two sampling strategies can be applied for testing the goodness of fit: the {\em scale-down}  and {\em back-in-time}, that is, the sample graph $S$ should either  have  similar (scaled-down) properties as compared to the original graph $G$, or  look like the original graph $G$  back at the time when it had the same size of $S$, respectively.
%Best performing sampling methods  for the scale-down sampling goal are those based on {\em random walks}, since they are biased towards high degree nodes and provide sampled graphs that are connected. For the back-in-time sampling goal, the {\em PageRank score-based} and  \emph{Forest Fire} type sampling  perform best~\cite{Leskovec:2006}.
\paragraph{Estimating Diameter by Probabilistic Counters.}
Palmer et al. proposed the \textsc{ANF} algorithm that exploits the (Flajolet-Martin) FM-counter~\cite{FlajoletMartin:1985} to derive the distance-based metrics of a graph. The core idea is to count the number of distinct nodes in each neighborhood $N(u,r)$, for all nodes $u$ and  radius $r$. For each set $N(u,r)$, \textsc{ANF} yields a concatenation of $l$ bit-masks (\emph{sketches}), where a bit-mask $l$ has probability $ \frac{1}{2^{i+1}}$ of having the $i$-th bit set to $1$.   %each with the probability of encountering $1$ at the $i$-th bit  that is equal to $ \frac{1}{2^{i+1}}$. The  $l$ bitmasks are built using  a family of independent hash functions.
%Although Palmer et al in ~\cite{palmer2002anf} do not show how the bitmasks are built, their  construction is described in the original paper by Flajolet and Martin~\cite{FlajoletMartin:1985}. 
An approximation of the number of distinct elements in a stream is derived by averaging the index of the least significant  bit with value $0$ 
in each of the $l$ bitmasks, and is set to $\frac{2^{\textrm{mean}}}{0.77351 }$~\cite{FlajoletMartin:1985}.
%If $h_l$ is the family of independent hash functions, %and $h_l(u)$ is  a 64-bit hashing value assigned to each node $u$, then 
%the algorithm computes for each node $u$ the longest sequence of $0$ values ending $h_l(u)$ (the tail of $h_l(u)$). The probability that $h_l(u)$ %the tail  of length $\tau(h(u))=i$ in binary representation  
%ends with the pattern $10^i$ is indeed $2^{-(i+1)}$. Tail lengths are aggregated within a bitmask  by the bitwise-OR operator, with $1$ appearing at the $(i+1)$-th position   with a probability equal to $ 2^{-(i+1)}$. An approximation of the number of distinct elements in a stream is derived by averaging the index of the least significant  bit with value $0$ 
%in each of the $l$ bitmasks, and is set to $\frac{2^{\textrm{mean}}}{0.77351 }$~\cite{FlajoletMartin:1985}.
Building upon this approach, Boldi et al. \cite{Boldi:2011:HAN} proposed \HB~ that uses the HyperLogLog algorithm \cite{DurandFlajolet2003,Flajolet07hyperloglog} and improves \textsc{ANF} in terms of speed and scalability,  providing a better estimate for the same amount of memory and number of passes.
%Alternatively to the original ANF algorithm, %instead of using a family of hash functions and  averaging these values, 
%different types of probabilistic counters may be applied, based on the LogLog and HyperLogLog algorithms~\cite{DurandFlajolet2003,Flajolet07hyperloglog}. Such algorithms are based on modifications of the original Martin-Flajolet algorithm to produce an improvement over ANF in terms of speed and scalability,  providing a better estimate for the same amount of memory and number of passes.
%In LogLog and HyperLogLog algorithms, a hashing function is used both to  uniformly assign all stream elements to $\m=2^b$ buckets (registers), based on  the first $b$ bits of their hash values: the remaining bits are used, within each bucket, to derive a maximum tail length, according to~\cite{FlajoletMartin:1985}. The estimate of the number of distinct elements in the stream is obtained by computing the mean of the maximum tail length over all the $\m$ buckets. LogLog and HyperLogLog use different approaches here: the former computes the arithmetic mean,  while the latter refers to the harmonic mean, which is more robust since it more heavily penalizes outliers.
%The   HyperLogLog estimate is defined as $R = \alpha_\m \m H$,
%where $ \alpha_\m$ is a given constant (approximately $1$), $H$ is the harmonic mean, and $\m$ is the number of  hash functions (sketches) applied. 
Although HyperLogLog is the best  approximate data stream counting algorithm, it is known that it tends to overestimate the real size of small sets~\cite{Heule2013:HPA}. Empirical bias correction has been introduced in~\cite{Heule2013:HPA}, where the correction works well in a good range of sizes, however errors persist on small sets where the \texttt{LinearCounting} algorithm \cite{Whang:1990:LPC}, provides the best results. %performs better than their bias-corrected raw estimate.  
%In fact, when the stream is small and the number of buckets is  approximately the number of elements in the stream, some  buckets may be empty and \texttt{LinearCounting} algorithm outperforms  HyperLogLog counters.
Alternatively, the MinHash technique can be used to estimate the size of the neighborhood with respect of All Distance Sketch (ADS) of a node of a weighted graph~\cite{Cohen:1997:SFA,Cohen:2014:ASR}. For each node, an ADS consisting of the first $ k $ MinHash is maintained. The estimate of the neighborhood of a node $u$ is given by hashing the nodes in the interval $[0,1]$ and filtering a node $v$ when its hash value is less than the $k$-th MinHash of the ADS, and when any other node in ADS is closer to $u$ than to $v$. %(is in the  ball $\Phi_{<u}(v)$).
This algorithm  computes, for each pair $(u,v)$ the {\em closeness similarity} which generalizes the inverse probability of the MinHash estimate~\cite{Bar-Yossef2002} with a Jaccard-like similarity function, that is $\frac{1}{max({\pi_{vx}} , {\pi_{ux}})}$ in the case of $k=1$, where $\pi_{u,v}$ is the Dijkstra rank of $u$ with respect to the node $v$ according to the position of $u$ by increasing distance from $v$.
% and with the expected size of an ADS of order $O(\log n)$ where $n$ is the size of the neighborhood, $\pi_{u,v}$ is the Dijkstra rank  of $u$ with respect to the node $v$ as the position of $u$ by increasing distance from $v$. 
%The estimate of the Dijkstra rank $\pi_{uv}$ of $u$ with respect to $v$ is  $1+\sum_{j\in ADS(v)\cap \Phi_{<u}(v)}\frac{k}{p_{vj}}$, being $p_{vj}$  the $k$-th value in the ADS list. 
If the graph is unweighted, then the BFS visit can be used and Cohen's framework can be considered equivalent in the spirit to HyperANF but with   the use of  the MinCount probabilistic counter  of~\cite{Bar-Yossef2002} instead of HyperLogLog probabilistic counter. 
%Cohen's  algorithm is similar to the first algorithm of~\cite{Bar-Yossef2002}.
However, its implementation is very different from the one presented in \cite{Bar-Yossef2002} and does not yield a $O(m)$ space complexity as in \cite{Bar-Yossef2002}. Amati et al. proposed a different probabilistic approach based on the \emph{MinHash} counter \cite{BIGDATA2017}, and experimentally showed its superiority in comparison to HyperLogLog based counters.  
Another sketching and sampling based technique to model public-private social network graphs proposes  to efficiently preprocess the public graph $G$ and to integrate it with a private user graph node in order to derive graph properties and measures \cite{Chierichetti_2015}. %A sketching-based algorithm is proposed to approximate the size of the reachability tree and the size of the private neighborhood at distance $d$ of a node $u$ to within $(1+\varepsilon)$-factor using $\bigO\left(m\cdot \Delta\cdot \frac{\log n}{\varepsilon^2}\right)$ preprocessing time, $\bigO\left(n\cdot \frac{\log n}{\varepsilon^2}\right)$ space and $\bigO\left(|E_u|\cdot \frac{\log n}{\varepsilon^2}\right)$ query time where $\Delta$ is the diameter of the public graph $G$ and $|E_u|$ is the size of the private edge set of node $u$. 

	\section{Preliminaries}\label{sec_preliminaris}
We proceed by formally introducing the terminology and concepts that we use in what follows. For $k\in \mathbb{N}$, we let $[k] = \{1,\dots ,k\}$.
%\paragraph{Graphs.}
%We start by introducing standard \emph{static} graphs\footnote{We use the terms ``graph'' and ``network'' interchangeably.}. 
An \emph{undirected graph}\footnote{We use the terms ``graph'' and ``network'' interchangeably.} is an ordered pair $G = (V, E)$, where $V$ is a set whose elements are called \emph{vertices} or \emph{nodes}, and $E$ is a set of \emph{unordered} pairs of vertices, whose elements are called \emph{edges}, or \emph{links} or \emph{arcs}. In a  \emph{directed graph} $G=(V, E)$, $E$ is a set of \emph{ordered} pairs of vertices.
%where $V$ is a set whose elements are called vertices or nodes, and $A$ is a set of ordered pairs of vertices, whose elements are called \emph{arcs}. 
Let $d(u,v)$ be the number of edges in the shortest path between $u$ and $v$. Given a graph $G=(V,E)$, define the \emph{neighborhood at distance at most $r$} for a node $u\in V$ as $N(u,r) = \{v\in V: d(u,v) \leq r\}$\footnote{Sometimes, we  use the term ``ball of radius $r$ centered in $u$'' to denote  $N(u,r)$.}. %Alternatively, $N(u,r)$ is the set of $r$-hop neighbors at distance at most $r$ from $u$
%We define the \emph{local neighborhood function} for a node $u\in V$ as the size of such set of nodes, i.e, $|N(u,r)|$. 
Additionally, we define the \emph{neighborhood function} at hop $r$ as the size of the set of pairs of nodes within distance $r$, formally: $|N(r)| =\left| \{(u,v)\in V\times V  : d(u,v)\leq r\}\right|$. %We define the distance metrics of our interest in terms of  the neighborhood function. 
The \emph{diameter} $\Delta$ of a graph is the longest shortest path in the graph. In terms of the neighborhood function we have:
%\begin{align}
    $\Delta = \min_{r\in [0,n-1]}\left\{r : \sum_{u}\left|N(u,r)\right| = \sum_{u}\left|N(u,r+1) \right| \right\}$.
%\end{align}
Similarly, the \emph{effective diameter} is defined as
%\begin{align}
    $\Delta^{\texttt{eff}} = \min_{r\in [0,n-1]}\left\{r : \sum_{u}\left|N(u,r)\right| \geq \uptau\cdot \sum_{u}\left|N(u,\Delta) \right| \right\}$
%\end{align}
for $\uptau \in [0,1]$. In this work, we consider $\uptau = 0.9$, i.e. the $90^{th}$ percentile distance between the nodes. 
We can also evaluate the \emph{average distance} of a graph $G=(V,E)$. Let $R(u,v)$ be the reachability function that assumes value $1$ if and only if $u$ can reach $v$ and $0$ otherwise. Thus we can write:
%\begin{align}
%\label{eq:avg_distance}
  $\texttt{AvgDist} = \frac{\sum_{u,v\in V}R(u,v)\cdot d(u,v)}{\sum_{u,v}R(u,v)}  =
  \frac{ \sum_{u}\sum_{r\in[\Delta]} \left(\left|N(u,r)\right|-\left|N(u,r-1)\right|\right)\cdot r}{\sum_u\left|N(u,\Delta)\right|}$.
%\end{align}
Observe that the \emph{number of reachable pairs} can be also defined using the neighborhood function as: $\texttt{Nr.Reachable Pairs} = \left|N(\Delta)\right|$.
Finally, we define the \emph{connectivity rate} $\alpha$ of a graph as the sparseness degree of its
transitive closure.
%\begin{align}
   $\alpha = \frac{1}{n\cdot (n-1)}\sum_{\substack{u,v\\ u\neq v}}R(u,v) \in [0,1]$.
%\end{align}
Notice that the more the graph is connected the higher is $\alpha$, and vice versa. As extreme values $\alpha = 1$ for a connected undirected graph, while $\alpha = 0$ when all the vertices are isolated.
\section{\Propagate\til Framework}\label{Sec:algorithm}
Any graph traversal algorithm, efficiently scans the edge list of a graph in a random order. However, if the algorithm needs to be efficient on graphs that do not fit in memory, we can not use standard graph traversal routines. As in \cite{palmer2002anf,Boldi:2011:HAN,BIGDATA2017}, we can find the nodes that are reachable from $u$ within $r$ hops by first retrieving their neighbors reachable in $r-1$ hops from $u$.
%The \Propagate\til framework is based on the observation that
%as \texttt{ANF} \cite{palmer2002anf}: 
Given $u$'s neighborhood at hop $0$, $N(u,0) = \{u\}$, we can compute $N(u,r)$ incrementally as: 
%the neighborhood $N(u,r+1)$ of radius $r$ around node $u$ can be derived as follows:
$N(u,r) = \bigcup _{(u,v)\in E}N(v,r-1)$.
%\begin{equation}
%\label{Eq:union}
%N(u,r) = \bigcup _{(u,v)\in E}N(v,r-1)
%\end{equation}
This technique allows to iterate over the edge set instead of performing a classical graph traversal. Probabilistic counters have been used to efficiently compute in terms of time and space the number of distinct elements in $N(u,r)$. The best known algorithms, namely \HB~\cite{Boldi:2011:HAN}, and \textsc{MHSE}~\cite{BIGDATA2017}, use respectively the HyperLogLog \cite{Flajolet07hyperloglog} and the MinHash counter and drop the required memory down to $2\cdot s\cdot n \cdot \log_2(\log_2 (n/s))$ bits and $2\cdot s\cdot n \cdot \log_2 n$ (where $s$ is the number of seed nodes from which we are starting the edge scan procedure). Even though their performance are impressive, they turn out to be prohibitive on very large graphs if our memory budget is low. Our novel framework overcomes such problems by using a clever implementation of a boolean array-like data structure allowing to have high-quality approximations of the distance-based metrics on machines with low memory requirements. Given a set of starting nodes $S = \{x_1,x_2,\dots,x_s\}$, \Propagate~
%As initial step, the \Propagate\til starts by drawing, from $V$, independently and uniformly at random (with replacement) a set
%$\{x_1,x_2,$
%$\ldots,x_s\}$ of $\m$ random nodes (seeds), and 
assigns to each node $u\in V$ a Boolean \emph{signature} array $\sigg{u}$ of length $s$ defined as follows: for all $ i\in[s]$ $\Sigg{u}{i} =\mathds{1}\left[u\in S\right]$, i.e., if the node $u$ is a seed, we set its coordinate to $1$.
%$\m$, defined as follows
%\[
%\Sigg{u}{i} = \left\{
%            \begin{array}{ll}
%            1 & \text{If } u = x_i\text{ is a seed} \\
%            0 & \text{ otherwise}
%            \end{array}
%            \right.
%\; \]
%the \emph{signature}, where $i$-th element is set to $1$ for the $s_i$ seed, $0$
%otherwise. 
%Furthermore, our algorithm exploits and improves the idea of the \mhse\til framework described in \cite{BIGDATA2017}. Given a set of $\m$ hash functions $H = \{h_1,\dots,h_s\}$ from a universal hashing family $\mathcal{H}$, each node $u\in V$ have an initial $\m$-signature $\sig{H}(\{u\})=\langle h_1(\{u\}),\ldots, h_\m(\{u\})\rangle$. 
%Starting from $\m$ initial seeds, 
Next, we extend the concept of signature to a \emph{set} of nodes of arbitrary size. Let $K\subseteq V$ be a subset of nodes, then its signature is defined as the \emph{bitwise} \texttt{OR} between the signatures of every node $u\in K$, formally $\Sigg{K}{i} = \bigvee_{u\in K}\Sigg{u}{i}$ for every $i\in [s]$. Notice that the $i^{th}$ index of  $\sigg{K}$ is equal to $1$ if and only if there exists at least one vertex $u\in K$ such that $\Sigg{u}{i}=1$.
%We extend the definition of signature to a set of nodes $N\subseteq V$, as $Sig_N[\;] = \bigvee_{u\in N}{Sig_u[\;]}$, that is the $i$-th index of  $Sig_N[\;]$ is set to $1$ if and only if there exists at least one vertex $u\in N$ such that $Sig_u[i]=1$.
The intuition behind our boolean signature is as follows. Suppose that we have only one seed node $x$, by definition its signature will be of the form $\sigg{x} = \langle 1\rangle$ i.e., $\Sigg{x}{1} = 1$. Subsequently, we expand the \emph{ball} centered in $x$ to its hop-1 neighborhood and for each neighbor $v\in N(x)$ we create a new signature $\Sigg{v}{\texttt{new}}$ equal to the bitwise \texttt{OR} between $\sigg{x}$ and $\sigg{v}$. After updating all $x$'s neighbors, for each $v\in N(x)$ we count the number of indices in its new signature that assumed value $1$ (one), we refer to the number of such indices as \emph{collisions} between the seed's bit and nodes' signatures. The total number of ones will be equal to the size of $x$'s hop-1 neighborhood. Observe that such value can be efficiently computed by summing the number of ones obtained by performing the \texttt{XOR} (exclusive \texttt{OR}) operation between $\sigg{v}$ and $\Sigg{v}{\texttt{new}}$ for each neighbor $v$, formally $|N(x)| = \sum_{u \in V} \lVert \sigg{u} \oplus \Sigg{u}{\texttt{new}} \rVert$. If we iterate this process $\Delta$ times, we will compute the number of nodes at distance of \emph{exactly} $r$ from $x$ for each $r\in [\Delta]$. By repeating this process $\Delta$ times for each node $x\in V$, we will obtain the \emph{exact} \emph{neighborhood function} $|N(r)|$ for each $r\in [\Delta]$. Recall that, under SETH, computing the exact neighborhood function cannot be done in $\bigO (n^{3-\varepsilon})$ for $\varepsilon>0$ \cite{Williams12}, thus we run the \Propagate~ framework on a subset of nodes $S$ sampled uniformly at random from $V$. Given a uniform sample of $s$ nodes from the vertex set $V$, the \Propagate\til framework can be implemented in two different ways: (1) every node has a signature of $s$ bits, and expands the $s$ balls (one for each seed node) in parallel until there is at least one signature that changed its value; (2) in a sequential fashion, every node spreads its bit until there is a signature that changed its value. We refer to these two implementations as \Bool~(Section ~\ref{Section:MHSE}), and \BoolSE~(Section \til\ref{Section:SE-MHSE}). \BoolSE\til is preferable  to \Bool\til when $s$ is very large and $s\cdot n$ bits becomes too big to be kept in the memory of a single machine. For example, when the set of seeds is the entire vertex set $V$, that is $\m=|V|$, then \Propagate~ computes the \emph{exact} neighborhood function. To compute the ground-truth values, \Bool~ needs $n^2$-bit array that, for big graphs, can be too large to be stored on a single machine. \BoolSE, instead,  needs only $n$ bits. Thus, for this task, \BoolSE\til is preferable to \Bool\til. In Section \ref{sec:exper_res} we compare the execution times of \BoolSE~  and the All Pair Shortest Path algorithm to compute the \emph{exact} neighborhood function of big real-world graphs.
%Note that the base-step of the \Propagate~ framework resembles a  BFS-like visit. However, due to the prohibitive dimensions of real-world networks we can not use a textbook breadth first search to propagate seeds in the graph. This is because in a multi-thread implementation (crucial to analyze huge networks) it is impossible to maintain the BFS queue. We opt for a clever ball-like expansion implementation that does not require auxiliary data structures to keep track of the visited vertices.

\subsection{\Bool\til Algorithm\label{Section:MHSE}}
%Based on all above observations, we can easily derive \bmhse\til %algorithm described by  Algorithm\til\ref{MHSE-Bool} which
%Based on all above observations, we can easily derive \Bool\til algorithm which
%implements the \Propagate\til framework.
%Instead of having $\m$ hash functions drawn from a family of hash functions  $\mathcal{H}$ as in \cite{BIGDATA2017}, we have 
Given $\m$ sample nodes $\{x_1, \dots, x_{\m}\}\subseteq V$, \Bool\til (Algorithm \ref{MHSE-Bool}) works as a synchronous diffusion process. It starts by initializing (line 2-3) the signature $\m$-array for each node $u\in V$, $\sigg{u}$ as described in Section~\ref{Sec:algorithm}. Subsequently, at each hop $r$, it computes for each node $u$ the \emph{signature} of the ball $N(u,r)= \{v\in V: d(u,v) \leq r\}$. The variable $\texttt{Count}$ (line 5) keeps track of the number of new collisions at hop $r$, that is the number of vertices at distance \emph{exactly} $r$ from $u$. The collisions at hop $r$ are subsequently stored in $\texttt{CountAll}[r]$ and if new collisions have been detected during the current hop, then the diameter lower-bound $\texttt{MaxHop}$ is updated, the approximated neighborhood function at hop $r$ is computed ($R[r]$ contains the number of pairs at distance at most $r$), the variable $\texttt{AvgDist}$ is increased with the difference between $R[r]$ and $R[r-1]$ times the hop $r$, and the hop $r+1$ is processed (lines 17-20).
Once the stopping criterion is met, i.e. no more collisions have been detected, the algorithm finds the minimum hop $r$ such that the ratio between the reachable pairs at hop $r$ and at the maximum hop is greater than $90\%$ i.e., computes the effective diameter $\Delta^{\texttt{eff}}$ (line 21), and normalizes the average distance value by dividing it with the maximum number of reachable pairs (line 22). Algorithm \Bool\til can be implemented using an array of $s$ bits for each vertex, thus we have the following theorem:
\begin{theorem}
	\label{th:Algo1}
	Algorithm \Bool\til (Algorithm\til\ref{MHSE-Bool}) computes the: diameter, effective diameter, average distance and number of reachable pairs in $\bigO(\Delta\cdot m)$ time using $\bigO(\m\cdot n+m)$ space.
\end{theorem}
\begin{proof}
	After the initialization of the node signatures, at every step of the \texttt{do-while}, Algorithm \til\ref{MHSE-Bool} scans the graph by iterating over the nodes. During iteration $r$, if a node $u\in V$ has at least one of its $s$ signature bits set to $0$, it updates its signature by performing a bitwise \textsc{OR} between its signature and the one of its neighbors. Notice that, once a bit flips to $1$, it cannot go back to $0$. Subsequently, for each node $u\in V$ it counts the number of ``bit-flips'' occurred in the current iteration, and stores the value in \texttt{CountAll}$[r]$. The algorithm stops when the set of bits that change values to $1$ is empty. Notice that a seed $s$ can perform at most $n-1$ hops. Since, at every iteration of the \texttt{do-while} every seed is propagated in parallel (as a synchronous diffusion process), the algorithm iterates for at most $\Delta$ steps. Every iteration of the \texttt{do-while} requires at most $\bigO(m)$ steps. Thus, the time complexity of the loop is $\bigO( \Delta \cdot m)$. Subsequently, the algorithm computes the effective diameter. This can be done in linear time in the diameter of the graph $\Delta$. Algorithm \til\ref{MHSE-Bool}, needs $s$ bits for each node signature, and two arrays of length $\Delta$ to store the number of collisions at each time step and the neighborhood function. Thus the space required by the algorithm is $\bigO\left(s\cdot n + m\right)$. The correctness of the algorithm follows from the definition of the \Propagate~ framework in Section \til\ref{Sec:algorithm}.
\end{proof}
%\inp
%\input{./camera_ready/PseudoCodice/MHSE-Bool_G}
%\input{PseudoCodice/MHSE-Bool_G}

\begin{algorithm}[htb!]
	%\scriptsize
    \KwData{$ G=(V,E)\; : |V| = n$, $\m$ sample of vertices $S\subseteq V$, eff. diameter threshold $\uptau$.}
	\KwResult{$\Delta^{\texttt{eff}}$ effective diameter, $\Delta_{\texttt{LB}}$ diameter, $R[\Delta_{\texttt{LB}}]$ number of reach. pairs, and $\texttt{AvgDist}$ average distance.}

    $ \Sigg{u}{i} = 0; \quad \forall u\in V, i\in [s]$ \tcp*{ $n\times s$ matrix of the nodes' signature}
 
 %${\Delta_{\texttt{LB}}} =  0$ %\tcp*{ \scriptsize keeps track of the diameter $\Delta$ }
	
	\For {\textbf{each }$x_i\in S$}{

	$\Sigg{x_i}{i} =  1$	%\tcp*{\scriptsize Initializaition of the seeds signatures. }
	}	
	$\texttt{CountAll}[0]= s,\texttt{Count} = 0,\texttt{AvgDist} =0,\texttt{r} =0, \Delta_{\texttt{LB}} = 0$
		
	$\texttt{R} = [0,0,\dots ,0]$ \tcp*{Neighborhood function}

	\tcp{ Process one hop at  a time for all the sample vertices $x_i$.}
	\Do{$\texttt{Count}>0$}{

	$\texttt{Count} = 0$ \tcp*{Collision counter for hop $r$}
	\ForEach{$u\in V$}{
	
	 %\nota{Questo if va riscritto usando l'operazione booleana effettuata nel codice che poi spieghiamo a parole nel paragrafo relativo a prop-P}
	 
	%\If{$\exists x_i $ \texttt{s.t. has not propagated trough $u$ yet}}{
		
		$\Sigg{u}{\texttt{next}} = \sigg{u}$
   	
        	\ForEach{ $\; u\rightarrow v$}{
        	 $\Sigg{u}{\texttt{next}} = \Sigg{u}{\texttt{next}}\vee\sigg{v}$ 
        	} 
	%}
	} 
	\ForEach{$u\in V$}{
	 $\texttt{Count = Count} + \lVert\Sigg{u}{\texttt{next}} \oplus\sigg{u}\rVert $	%\tcp{\scriptsize \textsc{count} is the 1's in the signatures \textsc{xor}.}
	
	 $\sigg{u} = \Sigg{u}{\texttt{next}}$\tcp*{Update $u$'s signature}
	}
	
	$\texttt{CountAll}[r] =  \texttt{Count}$ \tcp*{Reachable vertices at hop $r$}
	
	 ${\Delta_{\texttt{LB}}} = \max{\{\texttt{r},\Delta_{\texttt{LB}}\}}$\tcp*{Update diameter lower bound}
	 
	 $\texttt{R}[r] = R[r-1]+\texttt{CountAll}[r] $ \tcp*{$\texttt{R}[-1]$ treated as $0$ when $r = 0$}
	 
	  $\texttt{AvgDist = AvgDist} +r\cdot (\texttt{R}[r] -\texttt{R}[r-1]) $	\tcp*{$\texttt{R}[-1]$ treated as $0$ when $r = 0$}
	 
	 $\texttt{r = r} +1 $
	}
%	 $\texttt{R} = [0,0,\dots ,0]$ %\tcp{\scriptsize It contains the overall reachable pairs at distance $\leq r$}
%	 
%	\For{$k = 0$ \textbf{ to } $r$}{
%	 $\texttt{R}[k] = \left( n \cdot \sum_{i\leq k}{\texttt{CountAll}[i]}\right)/s $			
%	}

	%\For{$k = 0$ \textbf{ to } $r$}{

	% $\texttt{AvgDist = AvgDist} +k\cdot (\texttt{R}[k] -\texttt{R}[k-1]) $	\tcp*{$\texttt{R}[-1]$ treated as $0$ when $k = 0$}
	%}
	 
$\Delta^{\texttt{eff}} = \min_k \left\{k:\frac{R[k]}{R[\Delta_{\texttt{LB}}]}\geq \uptau\right\}$ \tcp*{Compute the effective diameter ${\Delta^{\texttt{eff}}}$}
	
  %${\Delta_{\texttt{LB}}} = \texttt{MaxHop}$ 	\tcp*{Lower bound for the diameter}
$\texttt{AvgDist = AvgDist} / \texttt{R} 
[\Delta_{\texttt{LB}}]$  \tcp*{ Compute the average distance }

	 $R[\Delta_{LB}] =  (n/s)\cdot R[\Delta_{LB}]$ \tcp*{ Compute the number of reachable pairs  }

	 \Return{$\Delta^{\texttt{eff}}$,$\Delta_{\texttt{LB}}$,$\texttt{R}[\Delta_{\texttt{LB}}],\texttt{AvgDist}$}
	\caption{\label{MHSE-Bool}\Bool\til Algorithm}

\end{algorithm}

\subsection{\BoolSE\til Algorithm\label{Section:SE-MHSE}}

%The sampling interpretation of \Propagate\til allows us to 
We derive an even more space efficient algorithm in which we process each sample vertex at a time using a single bit for each node in the graph, as with a Bernoulli process.  \BoolSE's pseudo code, is presented in the extended version of this paper \cite{Amati_2023}.
Differently from \Bool\til which maintains a signature $\m$-array for each vertex $u\in V$, \BoolSE\til uses a $n$-array $\sigg{V}$ that represents the \emph{signature} of the whole graph $G=(V,E)$. More precisely, given a seed node $x_i$ $\sigg{V}$, at each hop $r$, maintains the size of $x_i$'s neighborhood at distance at most $r$. Although, \BoolSE~has higher running time than \Bool, the independence of the seeds in \BoolSE~allows for a very simple implementation of the algorithm in a fully distributed and parallel processing, where cores or machines can be coupled with hash functions. Additionally, \BoolSE~can be implemented using \emph{progressive sampling} heuristics, that establish the sample size ``on the fly'' (see \cite{Amati_2023} for \BoolSE's incremental approach).  
%We only need one final reduction job to collect the whole number of collisions from each core or machine.
%Since, \sebmhse\til is easily distributable  we are currently working on a open source distributed and parallel  version of it  based on Apache Spark framework. This implementation is written in Scala and use Spark's Graphx  for managing the graphs. 
When $\m=|V|=n$, all \Propagate\til algorithms can compute the \emph{exact} distance-based metrics of interest. In this case, \Bool\til requires as signature a $n$ bit array for each vertex $u\in V$ thus requiring overall $n^2$ bits, which for large graphs is impracticable. However, \BoolSE\til would require only a $n$-bit array at each iteration and can be used to compute the \emph{exact} values for various graphs faster than the \texttt{APSP} algorithm implemented in \texttt{WebGraph} \cite{BoVWFI}\til(see Section\til\ref{Sec:tool}). 
For huge graphs, the only feasible algorithm in a standalone setting is the \BoolSE~algorithm.  
%and the space occupancy of \sebmhse\til would be  thus  necessary to accomplish standalone computations in memory for very large graphs. 
%In general, when $\m$ is very large,  \BoolSE\til is  the best algorithm  since it allows to reuse the arrays  among the $\m$ iterations. 
%On the other hand,  in real applications,  where good approximations are attained even for small values of $\m$ (see Theorems\til\ref{th:Algo1}, \til\ref{th:Algo2}	and\til\ref{th:MHSE-Error}),  then the choice between \Bool\til and  \BoolSE\til depends on whether we are in a standalone or a parallel (distributed) setting. 
The above  considerations lead to the following theorem:

% On the other hand,  when $\m$ is small, a distributed version of \sebmhse\til would be faster than \bmhse, so \sebmhse\til should be considered as a faster but parallel (distributed) version of \bmhse.  
% Equation (\ref{Eq:JaccardEstimation}) yields the Algorithm  \ref{al:MinHashingRealTime}, while Equation (\ref{Eq:JaccardEstimation2}) explains the Algorithm \ref{al:thirdVersion}. 
%Algorithm \ref{al:thirdVersion} has a very good  property  for reducing the space complexity. 
\begin{theorem}
	\label{th:Algo2}
	 \BoolSE\til computes the: diameter, effective diameter, average distance and number of reachable pairs in  $\bigO(\m\cdot\Delta\cdot m)$ time using  $\bigO(n+m)$ space.
\end{theorem}

\begin{proof}
	The proof is similar to that of Algorithm \til\ref{MHSE-Bool}. For each seed $s$, the algorithm scans the graph $\Delta$ times. Thus the time complexity is $\bigO(s\cdot \Delta\cdot m)$. In this case, the algorithm uses a signature of $n$ bits. Thus, the space complexity drops to $\bigO\left(n+m\right)$.
\end{proof}
\begin{algorithm}[htb!]
\scriptsize
		\KwData{
		$ G=(V,E)\; : |V| = n$, $\m$ sample of vertices $S\subseteq V$, eff. diameter threshold $\uptau$.}
		\KwResult{$\Delta^{\texttt{eff}}$ effective diameter, $\Delta_{\texttt{LB}}$ diameter, $R[\Delta_{\texttt{LB}}]$ number of reach. pairs, and $\texttt{AvgDist}$ average distance.}

      $ \Sigg{u}{}= 0; \quad \forall u\in V$ \tcp*{Graph' signature}

		$\texttt{CountAll}[0]= s,\texttt{Count} = 0,\texttt{AvgDist} =0,\texttt{r} =0, \Delta_{\texttt{LB}} = 0$
		
	    $\texttt{R} = [0,0,\dots ,0]$ \tcp*{Neighborhood function}
		
		 \tcp{Process each sample node $x_i$ at a time.}
		\ForEach{$x_i\in S$}{
		
		 \tcp{ Graph signature set to $1$ to the current sample node $x_i$ position, $0$ otherwise}
		 
		 	$\sigg{x_i} = 1$

			 $\texttt{Count} = 0$ \tcp*{ Initialize the collisions accumulator }

			$\texttt{r} = 0$
			
			\Do{$\texttt{Count}>0$}{			
			
				 $\texttt{NewSig = Sig}$
				 
				\ForEach{$u\in V$}{
				
%				
                   %\nota{Questo if va riscritto usando l'operazione booleana effettuata nel codice che poi spieghiamo a parole nel paragrafo relativo a prop-S}
                    
	                \If{$\sigg{u} = 0$}{
    					\ForEach{$u\rightarrow v$}{
    				
    							 $\texttt{NewSig(u) = NewSig(u)} \vee  \texttt{\sigg{v}}$
    							 
    							 \If{$\texttt{NewSig(u)} =1$}{
    							 
    							    \textbf{Break;}
    							 }
    					}	
    				}
			}
			
			 $\texttt{Count = Count} +\lVert \texttt{NewSig}\oplus\texttt{Sig}\rVert$ 		%	\tcp*{\scriptsize \textsc{count} is the $1$'s in the signatures \textsc{xor}.}

        ${\Delta_{\texttt{LB}}} = \max{\{\texttt{r},\Delta_{\texttt{LB}}\}}$\tcp*{Update diameter lower bound}			 
			 $\texttt{Sig} = \texttt{NewSig}$
			 
			 $\texttt{CountAll}[r] = \texttt{CountAll}[r]+\texttt{Count}$ %\tcp*{\scriptsize Reachable pairs at hop $r$}	
			 
			 $\texttt{r = r} + 1$
		} 
		}
		\tcp{ Compute distance based metrics.}
		$\texttt{R} = [0,0,\dots ,0]$ \tcp{ It contains the overall reachable pairs at distance $\leq r$}
		\For{$k = 0$ \textbf{ to } $r-1$}{
			$\texttt{R}[k] = \left( n \cdot \sum_{i\leq k}{\texttt{CountAll}[i]}\right)/s $			
		}
			\tcp{ Computation of the average distance}

		\For{$k = 0$ \textbf{ to } $r-1$}{
			
			$\texttt{AvgDist = AvgDist} +k\cdot (\texttt{R}[k] -\texttt{R}[k-1]) $\tcp*{ $\texttt{R}[-1]$ treated as $0$ when $k = 0$}
		}
		$\texttt{AvgDist = AvgDist} / R[\Delta_{\texttt{LB}}]$ \tcp*{Compute the average distance}

		$\Delta^{\texttt{eff}} = \min_k \left\{k:\frac{R[k]}{R[\Delta_{\texttt{LB}}]}\geq \uptau\right\}$
		\tcp*{ Compute the effective diameter ${\Delta^{\texttt{eff}}}$}
		
		 %\text{Same code of Algorithm \ref{MHSE-Bool}\til (lines 21-29)}
		 
		 %$\cdots\cdots$

		 \Return{$\Delta^{\texttt{eff}}$,$\Delta_{\texttt{LB}}$,$\texttt{R}[\Delta_{\texttt{LB}}],\texttt{AvgDist}$}

		\caption{\label{SE-MHSE}\BoolSE\til Algorithm}

	\end{algorithm}

%\subsection{Error Bounds of the sample size.\label{Section:ErrorBounding}}
\paragraph{Error Bounds of the sample size.\label{Section:ErrorBounding}}
We now
%Theorems\til\ref{th:Algo1}, \til\ref{th:Algo2} allows to 
evaluate the accuracy of the approximations of the \Propagate\til framework.
We use Hoeffding's inequality \cite{Hoeffding_1994} to obtain the sample size $\m$ for  good approximations  of the distance-based metrics of interest.

%We recall the definition of the \emph{reachability function} $R(u,u')$ as   $R(u,u')=1$ for all  reachable vertices $(u,u')$, $R(u,u')=0$, otherwise. Also, we define the \emph{connectivity rate} $\alpha$ as follows:
%\begin{equation}
%	\alpha=\frac{\sum_{u} \sum_{v}  R(u,v)}{n\cdot(n-1)}
%\end{equation}

%The connectivity rate measures the sparseness degree of the transitive closure graph, that is the more the graph is connected  the higher is $\alpha$, and vice versa. As extreme values, $\alpha=1$ for a connected undirected graph, while $\alpha=0$ when all the vertices are isolated.

\begin{theorem}
	\label{th:MHSE-Error}
	With  a sample of $\m=\Theta\left(\frac{\ln n}{\varepsilon^2}\right)$ nodes, with high probability (at least  $1-\frac{2}{n^2}$), \Propagate\til framework (\Bool\til and \BoolSE) compute:
	\begin{itemize}
		\item [i.]  the average distance with the absolute error bounded by $\varepsilon \frac{\Delta}{\alpha}$  
		\item [ii.] the effective diameter with the absolute error bounded by  $\frac{\varepsilon} {\tilde{\alpha}}$
		\item [iii.] the diameter with the absolute error bounded by $\frac{\varepsilon} {\tilde{\alpha}}$ 
		\item [iv.] the connectivity rate $\alpha$ with the absolute error bounded by $\varepsilon$ 
	\end{itemize}
where $\tilde{\alpha} = \alpha\cdot \frac{n-1}{n}$, and $\varepsilon>0$ a positive constant.
% \Propagate\til framework requires $\bigO(\frac{\ln n}{\varepsilon^2}\cdot m)$ time complexity; 
Thus, \BoolSE~requires $\bigO(\frac{\ln n}{\varepsilon^2}\cdot\Delta\cdot m)$ time and $\bigO(n+m)$ space . While, 
\Bool\til requires  $\bigO\left(n\frac{\log n}{\varepsilon^2}+m\right)$ space  complexity.
\end{theorem}
\begin{proof}
	
	Let us start with the \emph{average distance}. Let $X_i=\frac{n\cdot\sum_{u}  d(u,v_i)\cdot R(u,v_i)}{\sum_{u} \sum_{v}  R(u,v)}$ with $\{v_1,\dots, v_{\m}\} \subseteq V$ a sample, %a set of random variables defined as 
	%$X_i=\frac{\sum_{u}  d(v_i,u)\cdot R(v_i,u)}{n}$
	where  $X_i\in [0,\frac{\Delta}{\alpha}]$. The  expectation of $X_i$ is the  average distance:
	%of each $X_i$, that is:
	\begin{align}
		% E(X_i)={\sum_{u}X_i}*{P(u)}=\frac{\sum_{u}\sum_{v_i\in V} d(v_i,u)\cdot R(v_i,u)}{n^2}
		&\Expec(X_i)={\sum_{v_i\in V}X_i}\cdot{\Prob(v_i)}%= \frac{1}{n}\frac{ \sum_{v_i\in V}n\cdot  \sum_{u}d(u,v_i)\cdot R(u,v_i)}{ \sum_{u}\sum_{v_i\in V} R(u,v_i) }\\
		%	&
		=\frac{ \sum_{u}\sum_{v_i\in V} d(u,v_i)\cdot R(u,v_i)}{ \sum_{u}\sum_{v_i\in V} R(u,v_i) }
		\label{Eq:average_distance}
	\end{align}
	Then, Hoeffding's inequality \cite{Hoeffding_1994} generates the following bound:
	
	%	\begin{theorem}[Hoeffding's inequality]
		%Let $X_1,\dots, X_\m$ be independent bounded random variables with $a_i\leq X_i\leq b_i$,  where $-\infty <a_i\leq b_i<\infty$ for all $i$ and $\mu = \Expec\left(\sum_{i=1}^\m\frac{x_i}{\m}\right)$ is the expected mean. Then
		%
		%\begin{align}
		%\Prob\left(\left|\sum_{i=1}^\m{\frac{X_i}{\m}}-\mu \right|\geq \xi\right) \leq 2\exp{\left(-\frac{2\m^2\xi^2}{\sum_{i=1}^\m{\left(b_i-a_i\right)^2}}\right)}
		%\end{align}
		%for all $\xi\geq 0$.
		%\end{theorem} 
		\begin{equation}
			\Prob\left(\left|\frac{1}{\m} \cdot \sum_{i=1}^\m ( \Expec(X_i)- X_i) \right|\geq \xi\right)\leq 2\cdot\exp\left(\frac{-2\m\xi^2\alpha^2}{\Delta^2}\right)
			\label{Hoeffding}
		\end{equation} 
		If $\xi =\varepsilon \frac{\Delta}{\alpha}$ then Equation\til\ref{Hoeffding} becomes:
		
		\begin{equation}
			\Prob\left(\left|\frac{1}{\m} \cdot \sum_{i=1}^\m ( \Expec(X_i)- X_i) \right|\geq \varepsilon \frac{\Delta}{\alpha}\right)\leq 2\cdot\exp\left({-2\m\varepsilon^2}\right)
		\end{equation} 
		
		Therefore it is sufficient to take $\m=\Theta\left(\frac{\ln n}{\varepsilon^2}\right)$ to have, with high probability (at least   $1-2\cdot\exp\left({-2\ln n}\right)=1-\frac{2}{n^2}$),  an error at most $\varepsilon \frac{\Delta}{\alpha}$.
		
		In a similar way we can derive error bounds for the \emph{effective diameter}. Let $r$ be the effective diameter, and 
		
		\begin{equation}
			X^r_i=\frac{\sum_{\{u: d(u,v_i)\leq r\}}  n\cdot R(u,v_i)} { \sum_{u}\sum_{v} R(u,v) }=\frac{n\cdot|N(v_i,r)|}{|N(\Delta)|}
		\end{equation}
		where  $X_i^r\in [0,\frac{1}{\tilde{\alpha}}]$, where $\tilde{\alpha} = \alpha\cdot \frac{n-1}{n}$ . Again,  the expectation of the $X^r_i$ is: %$\sum_u {\frac{|N(u,r)|}{|N(\Delta)|}}=\frac{|N(r)|}{|N(\Delta)|}$.
		\begin{align}
			\Expec(X^r_i)= {\sum_{v_i\in V} X^r_i}\cdot{\Prob(v_i)}
			=\frac{\sum_{v_i\in V}  |N(v_i,r)| }{|N(\Delta)|}
			=\frac{ |N(r)| }{|N(\Delta)|}
			%	=\\\frac{1}{n}\displaystyle\frac{\sum_{\{u: d(u,v_i)\leq r\}} n \cdot 
				%	=	\sum_{i=1}^\m\frac {n\cdot R(v_i,u)}{\m}}{|\tau\cdot N(\Delta)|}  
		\end{align}
		%where each  $|N(v_i,r)|$ is  defined as on Equation \ref{Eq:JaccardEstimation2}. %\ref{Eq:DimN}.
		Applying  Hoeffding's inequality, with    $\xi =\frac{\varepsilon} {\tilde{\alpha}}$, we  approximate   $\frac{|N(r)|}{|N(\Delta)|}$   by
		%	$\frac{\sum_{i=1}^\m X^r_i}{s}=\displaystyle\frac{\sum_{\{u: d(u,v_i)\leq r\}} \sum_{i=1}^\m\frac {n\cdot R(u,v_i)}{\m}}{|N(\Delta)|}$
		$\frac{\sum_{i=1}^\m X^r_i}{s}=\displaystyle\frac{ n\cdot\sum_{i=1}^\m { |N(u,v_i)|}}{\m\cdot|N(\Delta)|}$
		with a sample of $\m=\frac{\ln n}{\varepsilon^2}$ nodes and an error bound of $\frac{\varepsilon} {\tilde{\alpha}}$, with high probability (at least $1-\frac{2}{n^2}$). Since the diameter $\Delta$ can be defined in terms of effective diameter $\min_{d'}\left\{d': \frac{ |N(d')|}{|N(\Delta)|}\geq \tau\right\}$ by choosing $\tau = 1$ instead of $0.9$, the effective diameter error bound holds also for the diameter. Finally, we give a bound for the number $|N(\Delta)|$ of \emph{reachable pairs}. Since $|N(\Delta)|$ is a very large number  we give the error bound for the ratio $\alpha= \frac{|N(\Delta)|}{n(n-1)}$. Let us define the random variable
		$
		X_i^\Delta = \frac{|N(v_i,\Delta)|}{n-1}
		$
		where $X_i^\Delta \in [0,1]$. The expected value of $X_i^\Delta$ is
		\begin{align}
			\Expec(X_i^\Delta) =%\sum_{v_i\in V}X_i^\Delta \cdot \Prob(v_i) =
			\sum_{v_i\in V}\frac{1}{n} \frac{|N(v_i,\Delta)|}{n-1}  = \frac{|N(\Delta)|}{n(n-1)}=\alpha
		\end{align}
		Applying Hoeffding's inequality % with $\xi = \varepsilon $ 
		we approximate $\alpha$ with  $\m=\frac{\log n}{\varepsilon^2}$ and error bound of $\varepsilon$, with high probability (at least $1-\frac{2}{n^2}$).
	\end{proof}

%Given the \emph{reachable pairs} error bound of Theorem \ref{th:MHSE-Error}.\textit{iv.} we have the following corollary.

%\begin{corollary}\label{cor:density_error}
%With a sample of $s=\Theta\left(\frac{\log n}{\varepsilon^2}\right)$ nodes, {\sc Propagate } framework computes the connectivity rate $\alpha$ with error bounded by $\frac{\varepsilon}{n}$.
%\end{corollary}
%Let us consider the neighborhood $N(v_i,D)$, $v_i$ a random sample of $\m$ nodes, and let $X_i=d(v_i,u)$ where $D$ is the diameter. Then the expectation of each $X_i$ on $N(v_i,D)$, that is $E(X_i) = {\sum_{u\in N(v_i,D)}X_i}*{P(u)}=\frac{\sum_{u\in N(v_i,D)}d(v_i,u)}{|N(v_i,D)|}$ is the average distance of the nodes of $N(v_i,D)$ from $v_i$.

In the following theorem we show that \Propagate~and \textsc{MHSE}~produce the same estimates. In other words, we can create a mapping between \Propagate's signature and \textsc{MHSE}'~signature. This implies that the results in Theorem~\ref{th:MHSE-Error} can be extended to the \textsc{MHSE}~Algorithm.
\begin{theorem}
	\label{MHSE-Embedding}
    Given a graph $G=(V,E)$ and a subset  $S\subseteq V$ of $s$ seeds, \Propagate~and \textsc{MHSE} produce the same set of reachable pairs $R[r]$ for $0\leq r\leq \Delta$.
\end{theorem}

\begin{proof}
It is sufficient to prove that both \textsc{MHSE} algorithm and \Propagate\til Algorithms produce the same set $R$ of of all reachable pairs at distance at most $r$. The key observation is that, for each collision with the minhash value of the graph for the \textsc{MHSE}~Algorithm we have a collision in the signature $\Sigg{\cdot}{}$ for the \Propagate\til Algorithm and viceversa.
Formally, let $\mathcal{H}$ be a set of $s$ hash functions over the nodes of the graph, and let $\Sigg{\cdot}{MHSE}$ be the signature with the minhash values of the nodes over the $\mathcal{H}$ hash functions. Let $\mathcal{U} = \{u_i\}$ be the set of nodes that have the minimum hash values with $i\in \mathcal{H}$. We define $\Sigg{u}{i} =1$ if and only if $u=u_i$. Viceversa, if we have a Boolean signature $\Sigg{\cdot}{}$ we may always  define a signature $\Sigg{\cdot}{MHSE}$ with $s$ hash functions having their minimum on the nodes with $\Sigg{u}{i} =1$.
At each hop and each hash function $i$, \textsc{MHSE}~counts the number of source nodes for which the merge operation of their signatures with the signature of their target nodes produces a new collision (see Amati et al.~\cite{Amati_2023} for more details), that is when $\Sigg{u}{MHSE}$ becomes the minimum, that is when $u=u_i$, that is when $\Sigg{u}{i} =1$, in other words, when there is a new collision for \Propagate. This show that the two Algorithms produce the same set of reachable pairs $R$ for each $0\leq r\leq \Delta$. Thus, they produce the same output. 
\end{proof}

	\section{Experimental Evaluation}
\label{Sec:tool}

In this section, we summarize the results of our experimental study on approximating the distance-based metrics in real-world networks. We compare our framework with the state-of-the-art algorithms to approximate the distance metrics, i.e., for each algorithm, we compute the \emph{average distance, effective diameter,} and \emph{number of reachable pairs}. Subsequently, we evaluate (using various metrics) how these estimates relates to the \emph{exact} ones computed by the All Pairs Shortest Path algorithm.

\subsection{Experimental Setting}\label{sec:exp_setting}

\paragraph{Algorithms.} Our study
includes several competitor algorithms for approximating the neighborhood function. We provide a short description and a space complexity analysis of the considered algorithms.
\begin{description}
    \item [\normalfont{\textsc{\HB:}}] The $\bigO( \Delta \cdot m)$ algorithm of Boldi et al. \cite{Boldi:2011:HAN,BoVWFI}, which uses HyperLogLog algorithms~\cite{DurandFlajolet2003,Flajolet07hyperloglog} to approximate the neighborhood function. \HB~ requires for each node $2^b = s$ registers that records the position $R$ with the bit $1$ starting the tail ending with all $0s$. More precisely, if $n$ is the number of distinct nodes in the graph, \HB~needs $2\cdot s\cdot n\cdot \log_2\left(\log_2\left(n/s\right)\right)$ bits for the registers.
    \item [\normalfont{\textsc{MHSE:}}] The $\bigO( \Delta \cdot m)$ algorithm of Amati et al. \cite{BIGDATA2017}, which uses the MinHash counter to approximate the neighborhood function. \textsc{MHSE} is based on a BFS visit and it requires an $\bigO(\log n)$ register for each node to record the signature, hence, it has the same space complexity as \textsc{ANF} (\textsc{ANF} maintains a bitwise $\bigO(\log n)$ register to count new incoming nodes in the stream, instead). \textsc{MHSE} requires $2\cdot s\cdot n \cdot \log_2 n$ bits.
    \item [\normalfont{rand-BFS:}] The algorithm by Eppstein and Wang \cite{Eppstein:2001:FAC}, which estimates the distance-based metrics using BFS visits starting from random nodes. Its time complexity is $\bigO(s\cdot m)$ and needs $\bigO(n+m)$ space.
    %\item [\normalfont{\Bool:}] The algorithm described in Section~\ref{Section:MHSE} for computing the distance based metrics. \Bool~ needs $n\cdot s$ bits because every node has a register of exactly $s$ bits. 
    %\item [\normalfont{\BoolSE:}] The algorithm described in Section~\ref{Section:SE-MHSE}.\BoolSE~ needs only $n$ bits ($1$ bit for every node in the graph). 
    \item [\normalfont{APSP:}] The Java implementation of the All Pair Shortest Path algorithm available in \texttt{WebGraph}\cite{BoVWFI}. The algorithm has been used to compute the exact values of the distance metrics and as a competitor algorithm for the second part of the experimental evaluation.
\end{description}

\paragraph{Networks}
\label{Sec:experimentation}
We evaluate all of the above competitors on real-world graphs of different nature, whose properties are summarized in Table \ref{tab::sparsityAnddatasetdim}. The networks come from two different domains: social networks and web-crawls. According to Theorem \ref{th:MHSE-Error}, the collection \texttt{BlackFriday}\footnote{The BlackFriday graph
is built from Twitter considering retweet and reply activities (\cite{IADIS2016}). This graph is comparable in size to the largest publicly available social network graphs, and  is very sparse.} should require larger number of samples than other collections,  because of a small connectivity rate  (see Table \til\ref{tab::sparsityAnddatasetdim}). %Additionally, \texttt{web-BerkStan} and \texttt{Orkut}, due to their long diameters,  should require a larger sample for the computation of the average distance. Indeed, these facts are experimentally shown on Figure \ref{fig:increment_undirected}, where a small number of samples may lead to poor results for such collections.

\begin{table}[htb!]
	\centering
%\scriptsize
\scalebox{1}{
	\ra{1.1}
	\begin{tabular}{lrrcccc}\toprule
	\textbf{Graph} & \textbf{n} &\textbf{m} & $\bm{\Delta}$ & $\bm{\alpha}$ & \textbf{Type} & \textbf{Ref.} \\\midrule
	%\texttt{Com-dblp}                          & 317080                    & 1049866                   & 23    & 1     & U             &      \cite{LAW}              \\
%\texttt{web-NotreDame}                      & 325729                    & 1497134                   & 93    & 0.168 & D             &         \cite{snapnets}           \\
%\texttt{Lasagne-Yahoo}                      & 653260                    & 2931708                   & 22    & 0.004 & D             &        \cite{konect}            \\
%\texttt{Com-Youtube}                        & 1134890                   & 2987624                   & 24    & 1     & U             &               \cite{snapnets}     \\
\texttt{BlackFriday}                        & 2700815                   & 3811922                   & 70    & 0.002 & D             &             \cite{IADIS2016}       \\
\texttt{Youtube-Links}                      & 1138495                   & 4942298                   & 23    & 0.446 & D             &       \cite{konect}            \\
%\texttt{Soc-pokec}                          & 1632803                   & 4942298                   & 18    & 0.8   & D             &         \cite{snapnets}           \\
%\texttt{Web-Google}                         & 875713                    & 5105039                   & 51    & 0.482 & D             &            \cite{snapnets}        \\
\texttt{Amazon-2008}                        & 7600595                   & 5158388                   & 48    & 0.854 & D             &               \cite{LAW}     \\
\texttt{Web-BerkStan}                      & 685230                    & 7600595                   & 715   & 0.488 & D             &             \cite{snapnets}       \\
\texttt{Twitch-Gamers}                      & 168114                    & 13595114                  & 8     & 1     & U             &              \cite{konect}      \\
\texttt{Hollywood-2009}                     & 1139905                   & 113891327                 & 12    & 0.88  & U             &             \cite{LAW}       \\
\texttt{Orkut-2007}                         & 3072441                   & 234370166                 & 61    & 0.356 & U             &               \cite{LAW}     \\\hline
\texttt{it-2004}                            & 41291594                  & 1150725436                & -     & -     & D             &                   \cite{LAW} \\
\texttt{gsh-2015-host}                      & 68660142                  & 1802747600                & -     & -     & D             &                \cite{LAW}    \\
\texttt{sk-2005}                            & 50636154                  & 1949412601                & -     & -     & D             &                 \cite{LAW}   \\
\texttt{gsh-2015}                           & 988490691                 & 33877399152               & -     & -     & D             &                 \cite{LAW}  \\
\texttt{clueweb12}                          & 978408098                 & 42574107469               & -     & -     & D             &                \cite{LAW}   \\
\texttt{uk-2014}                           & 787801471                 & 47614527250               & -     & -     & D             &                \cite{LAW}    \\
\texttt{eu-2015}                            & 1070557254                & 91792261600               & -     & -     & D             &  \cite{LAW} \\                
	
	\bottomrule
	\end{tabular}}
	
	\caption{The data sets used in our evaluation, where $n$ denotes the number of nodes, $m$ the number of edges, $\Delta$ the exact diameter, $\alpha$ the exact connectivity rate (type D stands for directed and U for undirected). The first seven graphs have been used in comparison of the four algorithms (\Propagate, \HB,\textsc{MHSE}, and \textsc{rand-BFS}) for accuracy and effectiveness, and speed. The last seven have been used to compare the performances of the algorithms on huge graphs. Dashed lines indicate that the exact metrics are not available due to the dimension of the data set.   	\label{tab::sparsityAnddatasetdim}}

	\vspace{-10mm}
\end{table}

\paragraph{Implementation and Evaluation details.} We released an open source platform for analyzing large graphs. This tool is developed in Java\footnote{\url{https://github.com/BigDataLaboratory/MHSE/tree/propagate-ecmlpkdd}} and uses some WebGraph libraries\til\cite{BoVWFI} to load and parse the  graph in compressed form. We chose WebGraph both for benchmarking with the compared algorithms and to allow us to: (1) compress very large graphs; (2) iterate the neighbor list of a node with faster random access; and, (3) use its offline methods to process very big graphs that cannot be loaded in memory. We executed the experiments on a server running Ubuntu 16.04.5 LTS equipped with AMD Opteron 6376 CPU (2.3GHz) for overall 32 cores and 64 GB of RAM. All the algorithms are fairly compared, i.e. using the same number of seeds/registers and cores. For the comparison between \Propagate, \HB, \textsc{MHSE}, and \textsc{rand-BFS} we use 256 sample nodes/registers and 32 cores. For the comparison between \BoolSE\til and APSP we use 32 cores. For the first part of the experiments, we repeat every test $10$ times and average over the results for every algorithm (\HB, \textsc{MHSE}, \textsc{rand-BFS}, \Bool, and \BoolSE). Whenever we are able to compute the \emph{{exact}} value $\hat{x}$ and thus the residual $(\hat{x}-\tilde{x})/\hat{x}$ where   $\tilde{x}$ its \emph{estimate} we also exhibit a p-value. More precisely, we perform a two-sided unpaired \emph{t-test} \cite{Case_Berg_2001} with confidence interval of $0.95$. Given a set $X$ of estimates of the distance metric $y$ obtained after $10$ runs of an algorithm $\mathcal{A}$, the null hypothesis is that its mean $\overline{X}$ is equal to the exact value $X$. If the displayed p-value is in the range $[0.9,1.0]$ then we fail to reject the null hypothesis, and conclude that the means are not significantly different. Therefore, we can conclude that algorithm $\mathcal{A}$ provides reliable and statistically significant estimates of $y$.
\subsection{Experimental Results}
\label{sec:exper_res}
%\paragraph{Results.}
%\subsection{Results}
\paragraph{Accuracy and effectiveness}
In our first experiment, we run on the networks listed in the first group of Table \ref{tab::sparsityAnddatasetdim} all the discussed approximation algorithms. 
%More precisely, on each data set, we run every algorithm $10$ times using $256$ registers/seeds. We then, for each of these four algorithms (\Bool,\BoolSE,\HB, plus \textsc{MHSE}): (1) compute the average of all the three estimated quantities (average distance, effective diameter, and number of reachable pairs); (2) evaluate the (average) precision (of every algorithm for every distance metric) by computing the residuals between the exact and the estimated values; and, (3) perform a two-sided unpaired \emph{t-test} between the vector of estimated values and the \emph{exact} distance metric of interest to evaluate the effectiveness of the algorithm.
In Table \ref{Experiment1}, we show the accuracy and effectiveness of all the competitor algorithms. \Bool~and \BoolSE~are grouped under the name of \Propagate, that is because both algorithm produce the same results. We observe that our novel framework leads the scoreboard against its competitors. It provides the best estimations in terms of accuracy and statistical significance. For the average distance, \Propagate~outperforms all the other algorithms on all the graphs except on \texttt{Orkut}, in which \textsc{rand-BFS} provides the best estimate. Moreover, it provides the best effective diameter estimates on  all the datasets. Finally, for the number of reachable pairs, \Propagate~provides very accurate estimations on all the networks except on \textsc{Youtube} for which \textsc{MHSE}'s estimate has lower residual.  Observe that \Propagate~ is the algorithm that provides the higher number of statistically significant estimations and does not perform worse than the competitor algorithms.

\paragraph{Speed.}
As a second experiment, we compared the average execution times of \Bool, \BoolSE, \HB, \textsc{MHSE}, and \textsc{rand-BFS}. In the left side of Table \ref{tab:estimation_times}, we show the running times (in milliseconds) of the algorithms. We observe that \Propagate~ framework outperform its competitors on almost every data set. Remarkably, \Bool, leads the scoreboard with the fastest execution times on four over seven graphs. It is slightly slower than \textsc{rand-BFS} on \texttt{balckFriday, Amazon-2008,} and \texttt{Web-BerkStan} i.e. the datasets with low connectivity rate and longest diameters for which a classic traversal algorithm should require less time than our framework. We point out that, \textsc{rand-BFS} does not scale well as the size of the graph increases (as shown in the next experiment). We observe that, on average, \Bool~ is $60\%$ faster than \HB~ and $81\%$ faster than \textsc{MHSE}. Moreover, \Bool~ outperform (in terms of speed) \HB, and \textsc{MHSE} on all the graphs. \HB's execution time is comparable with the one of \BoolSE~while \textsc{MHSE} is the slowest one. More precisely, \textsc{MHSE} is slower than every other algorithm on every network for which it does not require more than $64GB$ or \textsc{RAM} i.e., does not generate a memory overflow error.

\begin{table}[htb!]

	\centering
	%\scriptsize
	\ra{1.1}
	%\scalebox{1}{
\begin{tabular}{llllll||ll}
& \multicolumn{7}{c}{\textbf{Execution time}}  	\\ \cline{2-8}
& \multicolumn{5}{c}{\textbf{Milliseconds}}    &\multicolumn{2}{c}{\textbf{Hours}}                       \\ \cline{2-8} 
\textbf{Graph} & \textbf{Prop-P}         & \textbf{Prop-S}         & \textbf{\HB}    & \textbf{MHSE}    &\textbf{Rand-BFS} & \textbf{Prop-S} & \textbf{APSP}   \\\midrule

%\texttt{Com-dblp}                             & \textbf{102.484}  & 918.234          & 1286.4      & 1932.103  \\
%\texttt{web-NotreDame}                         & \textbf{408.181}   & 1928.478         & 4178.728 & 5919.012    \\
%\texttt{Lasagne-Yahoo}                         & \textbf{57.253}   & 662.588           & 1741.081  & 3633.200       \\
%\texttt{com-youtube}                           & \textbf{2498.716} & 9808.034        & 6544.003 & 10802.753  \\
\texttt{b.Friday}                           & 282.162    & 2899.075            & 22495.344 & 51034.047 & \textbf{69.21}&\textbf{9.513}    & 33.705 \\
\texttt{YT-Links}                         & \textbf{1761.891} & 5040.347        & 5986.410 & 6655.706  & 1771.25&\textbf{7.217}    & 11.118 \\
%\texttt{Soc-pokec}                             & \textbf{14252.906} & 15437.438          & 14425.925   & 32794.513   \\
%\texttt{Web-Google}                            & \textbf{4191.117} & 13041.319        & 5560.553 & 16426.488   \\
\texttt{Amazon}                           & 4339.072 & 12686.703         & 8451.259 & 195200.019 & \textbf{1767.12}&\textbf{11.027 }   & 11.914    \\
\texttt{W.Berk.}                          &       1535.781    & 2477.531  & 2741.563   & 5980.219  & \textbf{645.25} &\textbf{33.108}          & 122.59 \\
\texttt{Twitch-G.}                         &        \textbf{1562.219}  & 2901.497  & 2288.231  & 4486.044& 1863.10&\textbf{0.344}    & 3.617\\
\texttt{Hollywood}                        & \textbf{4113.060}     & 15537.897        & 11068.953 & 46409.728 & 7672.18&\textbf{47.77}    & 158 \\
\texttt{Orkut}                            & \textbf{2875.688}   & 8121.125            & 3833.688   &\quad\quad \xmark & 12783.13&\textbf{840}  & 960   
\end{tabular}
%}
\caption{For each network (column 1),  we show on the left side of the table the average execution time (in milliseconds) over ten runs for each algorithm. On the right side, we show the execution time (in hours) of \BoolSE, versus \textsc{WebGraph}'s APSP algorithm to compute the ground truth distance-based metrics. %The fastest algorithm is  highlighted in bold.
\xmark~ indicates that the experiment was interrupted due to a memory overflow error. \label{tab:estimation_times} }
\end{table}

\paragraph{Estimating Distance Metrics on huge graphs.}
As a third experiment, we run all the approximation algorithms on the biggest networks available in \cite{LAW} (see the second group of data sets in Table \ref{tab::sparsityAnddatasetdim}). We aim to to investigate the performances of all the competitor algorithms on very big graphs that cannot be loaded in the main memory. In Table \til\ref{tab::stress_test}, we show the running times of the approximation algorithms. The first column indicates whether the graph can be fully loaded in memory in its uncompressed form. If this is not possible, we use \texttt{WebGraph}'s offline methods to access the compressed graph from the disk without loading it in memory. Observe that accessing the compressed graph directly from the disk, slows down the overall execution of the algorithms. However, it is the only way to analyze these graphs with our $64$GB memory machines. We observe that \BoolSE~can compute the distance metrics on \emph{every} graph. Considering the size of the data sets, \BoolSE~requires a reasonable amount of time to approximate the neighborhood function using $256$ seeds. \Bool~can compute the distance metrics for \texttt{it-2004, gsh-2015-host}, and \texttt{sk-2005}. Moreover, it is still possible to run \Bool~on the remaining graphs by appropriately decreasing the number of sample nodes. Instead, \HB~can be used only to compute the approximated neighborhood function only on \texttt{it-2004}, and \texttt{sk-2005}. Finally, \textsc{MHSE} and \textsc{rand-BFS} cannot be used on any of these networks. Before comparing the time performances, we point out that the red dash ({\color{red} --}) in Table \ref{tab::stress_test} indicates that the algorithm requires more than $64$GB of memory even with $1$ seed/register. Thus, decreasing the number of seeds/registers is not enough to run these algorithms on these huge networks, we would need to upgrade the RAM of the machine. From the results in Table \ref{tab::stress_test}, we observe that (on \texttt{it-2004}, and \texttt{sk-2005}) \Bool~is on average $64\%$ faster than \HB. Furthermore, \BoolSE~is the best algorithm to approximate the neighborhood function on huge data sets. It requires $n\cdot s$ bits, to store the graph \emph{signature}. Indeed, for \texttt{eu-2015}, i.e. the biggest graph in Table \ref{tab::sparsityAnddatasetdim}, it needs approximately at most $1$GB to store the graph signature. Thus, using \texttt{WebGraph}'s offline methods to scan the graph, \BoolSE~could provide the approximated neighborhood function of \texttt{eu-2015} using an average \emph{laptop}.

\begin{table}[htb!]

	\centering
	%\scriptsize
	\ra{1.1}
	%\scalebox{1}{
	\begin{tabular}{@{}lcccccc@{}}
&	& \multicolumn{4}{c}{\textbf{Execution Time}}                        \\ \cline{3-7} 
	\textbf{Graph} & \textbf{In memory} & \textbf{Propagate-P} & \textbf{Propagate-S} & \textbf{HyperANF} & \textbf{MHSE} &\textbf{rand-BFS} \\\midrule
	\texttt{it-2004}        & Yes                & 33.26 minutes        & 52.13 minutes        & 62.18 minutes     & {\color{red}--}  & {\color{red}--}     \\
	\texttt{gsh-15-h}  & No                 & 40 minutes           & 4.16 hours           & {\color{red}--}                &{\color{red}--}   &{\color{red}--}           \\
	\texttt{sk-2005}        & Yes                & 44 minutes          & 6 hours              & 4 hours           & {\color{red}--}   & {\color{red}--}            \\
	\texttt{gsh-2015}      & No                 & \xmark                    & 11 hours             & {\color{red}--}                 & {\color{red}--}    & {\color{red}--}           \\
	\texttt{clueweb12}      & No                 & \xmark                    & 9.56 hours           & {\color{red}--}                 & {\color{red}--}    & {\color{red}--}          \\
	\texttt{uk-2014}        & No                 & \xmark                   & 39 hours             & {\color{red}--}                 & {\color{red}--}    & {\color{red}--}         \\
	\texttt{eu-2015}        & No                 & \xmark                   & 7 days               & {\color{red}--}               & {\color{red}--}      & {\color{red}--}       \\
\bottomrule
\end{tabular}
%}
\caption{For each network (column 1), we show the loading method (column 2) ``Yes'' means that it is possible to load the entire graph in memory, while ``No'' indicates that is not possible. In such a case, we use \texttt{WebGraph}'s offline methods to iterate trough the successor lists. For each algorithm, we show the execution time (using $256$ seeds/registers). \xmark~ indicates that the experiment was interrupted due to a memory overflow error of the algorithm while initializing the signature/registers array. Here, the red dash {\color{red}--}~indicates that the algorithm cannot run even with $1$ seed/register.
 \label{tab::stress_test}}

\end{table}

\paragraph{Computing ground truth metrics with \Propagate.}
As a last experiment, we compare \BoolSE~with the \texttt{WebGraph} implementation of the All Pair Shortest Path (\textsc{APSP}) algorithm to compute the \emph{exact} neighborhood function. Among all the competitor algorithms, \HB~cannot be used to compute the ground truth values of a graph. That is because its neighborhood function estimator that uses the \texttt{HyperLogLog} counter is \emph{asymptotically almost unbiased} \cite{Boldi:2011:HAN}. \textsc{MHSE} instead, cannot be employed because of its high space complexity. Observe that \textsc{rand-BFS} coincides with \texttt{WebGraph}'s APSP algorithm. As showed in the proof of Theorem \ref{th:MHSE-Error} (see \cite{Amati_2023}) \Propagate's distance metrics estimators are all \emph{unbiased}. Thus, our novel framework can be used to compute the ground truth values. Given a $n$ vertices graph $G=(V,E)$, \Propagate~suffices of the entire vertex set $V$ as set of seeds to compute the exact distance metrics. For this experiment, we use \BoolSE~because it needs only $n$ bits to store the graph signature, while \Bool~would need $n^2$ bits and with a $64$GB machine can be used only for computing the ground truth metrics of the first three graphs in Table \ref{tab::sparsityAnddatasetdim}. In the right side of Table\til\ref{tab:estimation_times}, we show the running times of \BoolSE~and \textsc{APSP}. We observe that \BoolSE~is faster than \texttt{WebGraph}'s \textsc{APSP} implementation on all the data sets. %Overall, our framework is much faster than the \texttt{APSP} algorithm.\BoolSE~ shows worse performances on few data sets, such as: \texttt{web-NotreDame, web-Google,} and \texttt{web-BerkStan}.
These results suggest that our implementation of \BoolSE~is preferable for retrieving exact values of distance-based metrics on very large real-world graphs.

\begin{figure}[t!]
\centering
	
	\begin{subfigure}[b]{0.3\textwidth}
	
			\includegraphics[scale=0.45]{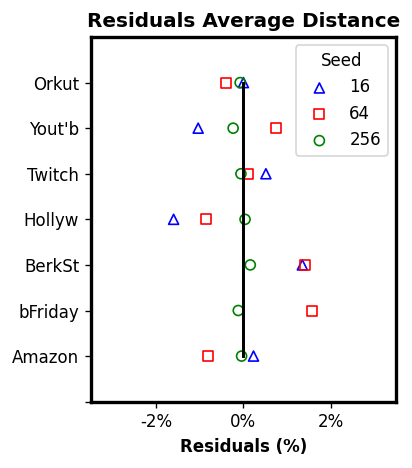}
			\caption{}
	
		\end{subfigure}
	\begin{subfigure}[b]{0.25\textwidth}
			
			\includegraphics[scale=0.45]{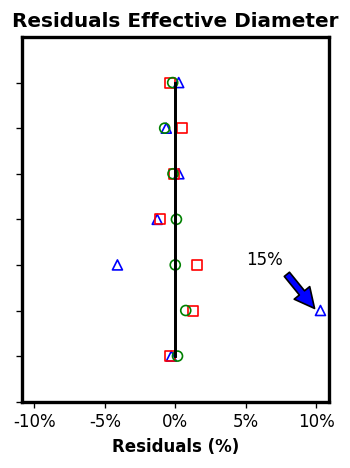}
			\caption{}
	
		\end{subfigure}
	\begin{subfigure}[b]{0.3\textwidth}
			
			\includegraphics[scale=0.45]{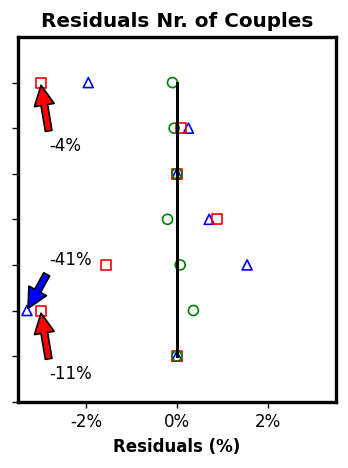}
			\caption{}
	
		\end{subfigure}

	\caption{Effectiveness of \Propagate\til on real world graphs with different number of seeds. Plots of the residuals for: Average Distance, Effective Diameter at 90\% , and Number of Connected Pairs.
			\label{fig:increment_undirected}}
\end{figure}

\paragraph{\BoolSE's incremental approach.}
In this experiment, we aim to better investigate the dependency of \Propagate's performance with respect to its sample size. We recall that Theorem \ref{th:MHSE-Error} states that a sample of $\bigO(\log n)$ nodes is enough to obtain good approximations of the average distance and the effective diameter with high probability, although $256$ seeds give really good approximations (see Table \til\ref{Experiment1}). Figure \ref{fig:increment_undirected}, shows the performances of the \Propagate\til framework with a representative increasing number of seeds, that is  $\m=16, 64, 256$. \Propagate\til achieves high quality approximations with $64$ seeds even for \texttt{Orkut-2007} and \texttt{Web-BerkStan} i.e. the graphs that have low connectivity rate $\alpha$. Our novel framework performs well for each approximation (average distance, effective diameter, and number of reachable pairs) with $64$ seeds on both directed and undirected graphs.
 Figure \ref{fig:increment_undirected} (c) shows the effectiveness of the framework as a probabilistic counter of the set of reachable nodes for both, directed and undirected graphs. Therefore, \BoolSE\til can be used with a progressive sampling approach \cite{Riondato_2015}. Indeed we could start with a small sample of nodes and progressively increase it if a certain stopping criterion is not met yet, allowing for an even faster approximation approach.

\section{Conclusions}
\label{Sec:conclusion}
We proposed \Propagate, a novel framework for estimating distance-based metrics on very large graphs. In Section \ref{Sec:algorithm}, we provided two different implementation of our framework, that, so far, can approximate: average distance, (effective) diameter, and the connectivity rate up to a small error with high probability. Our experimental results are summarized in Section \ref{sec:exp_setting}, which depicts the performance of our framework versus the state-of-the-art algorithms. Our approach over-perform in terms of accuracy and running time all its competitors. Moreover, when applied to very large real-world graphs, \BoolSE~(and \Bool~if applicable) clearly outperforms all the other algorithms in terms of scalability. As indicated in Table \ref{tab::stress_test}, our framework is the only available option to approximate distance-based metrics when we do not have access to servers with a large amount of memory. In the spirit of reproducibility, we developed an open source framework in Java that allows any user with an \emph{average laptop} to approximate the distance-based metrics considered in this paper on any kind of graph. Some promising future directions are to use \Propagate~to compute centrality measures on vertices and edges, and to extend our framework to community detection tasks.

\subsubsection*{Acknowledgements.} This work was partially supported by the
European Union under the Italian National Recovery and Resilience Plan (NRRP) of
NextGenerationEU, partnership on ``Telecommunications of the Future'' (PE00000001 program ``RESTART'')

\begin{table}[htb!]
	\centering
	
	\ra{1}
	\begin{tabular}{ccccccccl}
		&                          & \multicolumn{6}{c}{\textbf{Neighborhood Function Estimation}}                                                                                                                                                                                                                                                                                                                                                       &                      \\ \cline{3-8}
		&                          &                                                              &                                                                    &                                                                      & \multicolumn{3}{c}{\textbf{Residual/$\hat{x}$ (p-value)}}                                                                                                                                                      &                      \\ \cline{6-8}
		\rotatebox[origin=c]{90}{\textbf{Graph}}                                               & \textbf{Algo.}          & \textbf{\begin{tabular}[c]{@{}c@{}}Av.\\ Dist.\end{tabular}} & \textbf{\begin{tabular}[c]{@{}c@{}}Eff.\\ Diam.\\ 90\end{tabular}} & \textbf{\begin{tabular}[c]{@{}c@{}}Nr. of conn\\ pairs\end{tabular}} & \textbf{\begin{tabular}[c]{@{}c@{}}Av.\\ Dist.\end{tabular}} & \textbf{\begin{tabular}[c]{@{}c@{}}Eff.\\ Diam.\\ 90\end{tabular}} & \textbf{\begin{tabular}[c]{@{}c@{}}Nr. of conn\\ pairs\end{tabular}} &                      \\ \cline{1-8}

		\multirow{3}{*}{\rotatebox[origin=r]{90}{\texttt{blackFriday}}} & \texttt{Exact}$(\hat{x})$ & 16.124                                                       & 22.722                                                              & 11,300,563,035                                                       &                                                              &                                                                    &                                                                      &                      \\
		&\texttt{ \PropagateS}\til                     & 16.143                                                       & 22.551                                                             &
		11,259,575,354                                               & \textbf{-0.001(0.92$^\bullet$)}                                      & \textbf{-0.008(0.95$^\bullet$)}                                             & \textbf{0.003(0.72)}                                                          &                      \\
		& \texttt{H.ANF}$^{(\blacktriangle)}$                & 16.214                                                       & 22.841                                                             & 11,032,542,659                                                       & 0.01(0.26)                                                  & 0.005(0.40)                                                        & -0.024(0.34)                                                         &                     \\ & \texttt{MHSE}				& 16.338&23.029&12,193,068,803 &-0.01(0.12)&-0.01(0.23) &-0.07(0.09)\\
		&\texttt{rnd-BFS}				& 17.381&24.30&8,636,102,833 &-0.078(0.60)&-0.069(0.68) &0.24(0.09)
		\\\cline{1-8}		
		
		\multirow{3}{*}{\rotatebox[origin=r]{90}{\texttt{Youtube}}}                               & \texttt{Exact}$(\hat{x})$ & 5.104                                                        & 6.244                                                              & 577,863,455,179                                                      &                                                              &                                                                    &                                                                      &                      \\
		& \texttt{\PropagateS}\til                     & 5.104                                                        & 6.291                                                                & 578,216,139,787                                                     &  \textbf{0(1$^{**}$)}                                                 & \textbf{-0.007 (0.44)}                                                       & -6e-4\textbf{ (0.73)}                                                         &                      \\
		& \texttt{H.ANF}                & 5.131                                                        & 6.301                                                              & 602,314,527,291                                                      & -0.005 (0.1)                                                 & -0.009 (0.11)                                                      & -0.042 (0.1)                                                         &                      \\ & \texttt{MHSE}				&5.105 &6.165& 577,569,359,888&0.003(0.25)&0.013(0.08)& \textbf{5e-4}(0.65)&\\
		& \texttt{rnd-BFS}				&5.11&	6.217&	579,121,217,963&	-0.001(0.72)&	0.004(0.60)&	-0.002(0.01)
		\\\cline{1-8}

		\multirow{4}{*}{\rotatebox[origin=r]{90}{\texttt{Amazon}}}                                 & \texttt{Exact}$(\hat{x})$ & 12.075                                                       & 15.544                                                             & 461,523,315,650                                                      &                                                              &                                                                    &                                                                      &                      \\
		& \texttt{\PropagateS}\til                     & 12.08                                                        & 15.519                                                             &  461,523,315,650                                                         & \textbf{0.00(0.93$^\bullet$)}                                         & -0.002(0.80)                                                       & \textbf{0.00(1$^{**}$)}                                                 &                      \\
		& \texttt{H.ANF}               & 12.042                                                       & 15.47                                                              & 451,448,606,322                                                      & -0.003(0.44)                                                 & -0.022(0.31)                                                       & -0.022(0.31)                                                         &                      \\ & \texttt{MHSE}				&12.1 &15.542& 462,552,552,254& -0.002(0.54)& \textbf{1e-4(0.98$^*$)}&-0.002(0.65) &\\
		&\texttt{rnd-BFS} &            12.103                                           &       15.579                                                      &  461,522,729,233                                                &                         -0.002(0.36)                    &                                               -0.002(0.53)         &                                                 1.27e-6(0.05)      &                     
		\\\cline{1-8}

		\multirow{3}{*}{\rotatebox[origin=r]{90}{\texttt{BerkStan}}}                                & \texttt{Exact}$(\hat{x})$ & 13.905                                                       & 17.777                                                             & 229,179,533,137                                                      &                                                              &                                                                    &                                                                      &                      \\
		& \texttt{\PropagateS}\til                     & 13.883                                                        & 17.777                                                             & 229,015,123,311                                                      & 	\textbf{0.02}(0.42)                                                 & \textbf{0(1$^{**}$ )}                                             & \textbf{7e-4}(0.65)                                                    & \multicolumn{1}{c}{} \\
		& \texttt{H.ANF}                 & 14.645                                                       & 17.728                                                             & 233,108,112,819                                                      & 0.053(0.40)                                                  & -0.003(0.83)                                                       & 0.017(0.49)                                                          & \multicolumn{1}{c}{} \\ 
		& \texttt{MHSE}				& 15.29& 18.14& 239,485,315,387&0.099\textbf{(0.61)}&0.02(0.61) &0.045(0.36)\\
		& \texttt{rnd-BFS}				& 14.341& 18.02& 228,188,757,612&-0.031(0.44)&-0.02(0.28) &0.004\textbf{(0.80)}\\\cline{1-8}
		
		\multirow{3}{*}{\rotatebox[origin=r]{90}{\texttt{{Twitch}}}} & \texttt{Exact}$(\hat{x})$ & 2.876                                                        & 3.127                                                               & 28,262,316,996                                                        &                                                              &                                                                     &                                                                       \\
		& \texttt{\PropagateS}\til                     & 2.876                                                        & 3.129                                                               & 28,262,316,996                                                        &  \textbf{0.0(1$^{**}$)}                                         & \textbf{-0.001 (0.94$^\bullet$)}                                             & \textbf{0.00(1.00$^{**}$)}                                                     \\
		& \texttt{H.ANF}                & 2.891                                                        & 3.180                                                               & 28,451,734,342                                                        & -0.005 (0.03)                                                & -0.017 (0.06)                                                       & -0.007 (0.78)                                                         \\ 
		& \texttt{MHSE}				&2.881 &3.140&28,262,316,996&-0.002(0.44)&\textbf{-0.001}(0.62) &\textbf{0.00(1.00$^{**}$)}     \\
		&\texttt{rnd-BFS}		&	2.868&	3.096&	28,262,020,235		&0.0027(0.33)&	0.01(0.29)	&1e-5(0.01)
		  \\\cline{1-8}

		\multirow{3}{*}{\rotatebox[origin=r]{90}{\texttt{Hollywood}}} & \texttt{Exact}$(\hat{x})$ & 3.855                                                        & 4.394                                                               & 1,143,030,619,175                                                     &                                                              &                                                                     &                                                                       \\
		& \texttt{\PropagateS}\til                     & 3.855                                                       & 4.397                                                               & 1,143,485,513,294                                                   & \textbf{-2e-4(0.92$^\bullet$)}                                                   & \textbf{-3.9e-4(0.95$^{*}$)}                                                          & \textbf{-8e-4(0.90$^\bullet$)}                                                           \\
		&\texttt{H.ANF}              & 3.848                                                        & 4.374                                                               & 1,136,104,164,355                                                     & 0.16(0.49)                                                   & -0.46(0.5)                                                          & 0.61(0.80)      
		\\ 
		& \texttt{MHSE}				&3.857 & 4.382 & 1,138,248,927,314& -3.91e-4 (0.86) &0.003 (0.65)&  0.0042(0.56)\\ 
		&\texttt{rnd-BFS}		&	3.840&	4.364&	1,143,960,403,951&	0.004(0.18)&	0.007(0.16)&	-0.001\textbf{(0.90$^\bullet$)}
		 \\\cline{1-8}
		\multirow{3}{*}{\rotatebox[origin=r]{90}{\texttt{Orkut}}} & \texttt{Exact}$(\hat{x})$ & 6.397                                                       & 8.906                                                             & 3,359,893,990,935                                                    &                                                              &                                                                     &                                                                       \\
		& \texttt{\PropagateS}\til                     &  6.402                                                      &8.895                                                             & 3,386,716,107,589                                                     & {-7e-4(0.90$^{\bullet}$)}                                                   & \textbf{0.001(0.82)}                                                          & \textbf{-0.008(0.60)}                                                           \\
		&\texttt{H.ANF}              & 6.389                                                      &8.917                                                            &  3,336,311,224,217                                                   & 0.001(0.53)                                                   & \textbf{-0.001}(0.76)                                                          & 0.01(0.45) \\                                                                                                               
		& \texttt{MHSE}				&\xmark&\xmark& \xmark& \xmark&\xmark&  \xmark \\
		& \texttt{rnd-BFS}		&6.399&	8.929&	3,328,105,314,542&	\textbf{-4e-4(0.91$^{\bullet}$)}&	-0.003(0.61)&	0.01(0.32) \\\cline{1-8}

	\end{tabular}
	\caption{The comparison of HyperANF,  \Propagate, and MHSE  using 10 trials and 256 registers and 256 sample nodes respectively. Statistical significance at the $90\%$, $95\%$ and $99\%$ confidence level are marked with $^\bullet $, $\textbf *$ and $\textbf {**}$ respectively. The algorithms requiring more heap size are marked with $^\blacktriangle$ , and \xmark~ indicates that the algorithm needs more than $64$GB of memory. \label{Experiment1} }
\end{table}

	\clearpage

	\bibliographystyle{splncs04}
	\bibliography{bigdata}
	%
	
	%\newpage
	%\clearpage
	%\appendix
	%\begin{center}
		%\LARGE{\textbf{Additional Materials}}
	%\end{center}
	%\input{../trunk/appendix}
	%\input{./trunk/appendixRevised}
	
\end{document}